\documentclass[12pt]{article}
\usepackage[english]{babel}
\usepackage{epsfig}
\usepackage{rotate}

\def\pin{$\mbox{}$\indent}  

\def\bbt#1{\bibitem{#1} \label{bb:#1}}

\newcounter{shimeqsno} 

\def\bsigma{\mbox{\boldmath $\sigma$}}
\def\bxi{\mbox{\boldmath $\xi$}}

\def\sign{\mbox{\rm sign}}

\def\E{\mbox{\rm E}}
\def\Var{\mbox{\rm Var}}
\newcommand  {\Rbar} {{\mbox{\rm$\mbox{I}\!\mbox{R}$}}}
\def\Cov{\mbox{\rm Cov}}

\def\dN{{\cal N}}

\def\ustr#1#2{\;\,\stackrel{#1}{#2}\;\,}

\let\Journal=\it

\def\bc#1{{\Journal Biol. Cybern.} {\bf #1}}
\def\el#1{{\Journal Europhys. Lett.} {\bf #1}}

\def\jpa#1{{\Journal J. Phys. A: Math. Gen.} {\bf #1}}

\def\jpp#1{{\Journal J. Physique} {\bf #1}}

\def\jsp#1{{\Journal J. Stat. Phys.} {\bf #1}}

\def\phc#1#2{{\Journal Physica #1} {\bf #2}}

\def\pra#1{{\Journal Phys. Rev. A} {\bf #1}}
\def\pre#1{{\Journal Phys. Rev. E} {\bf #1}}

\def\zpb#1{{\Journal Z. Phys. B} {\bf #1}}

\begin{document}
\setcounter{page}{0}
\begin{center}
{\huge Parallel dynamics of fully connected 
${\mbox{\boldmath Q}}$-Ising neural networks}
\\ \vspace{1.cm}
{\large 
D.~Boll\'e \footnote{Also at Interdisciplinair Centrum voor Neurale
	Netwerken, K.U.Leuven}
	\footnote{e-mail: desire.bolle@fys.kuleuven.ac.be}
        and G.~Jongen $^1$ \footnote{e-mail: greetje.jongen@fys.kuleuven.ac.be}
	\\Instituut voor Theoretische Fysica,
	  K.U.\ Leuven \\
	  B-3001 Leuven, Belgium \\
        \vspace{0.5cm}
        and G.~M.~Shim \footnote{ e-mail: gmshim@nsphys.chungnam.ac.kr}
        \\Department of Physics, Chungnam National University \\
          Yuseong, Taejon 305-764, R.O.~Korea}
\end{center}
\date{}
\thispagestyle{empty}
\vspace{0.5cm}
\begin{abstract}
\normalsize
\noindent
Using a probabilistic approach we study the parallel dynamics of fully
connected $Q$-Ising neural networks for arbitrary $Q$. A Lyapunov function
is shown to exist at zero temperature. A recursive scheme is set up to
determine the time evolution of the order parameters through the evolution
of the distribution of the local field. As an illustrative example, an
explicit analysis is carried out for the first three time steps.
For the case of the $Q=3$ model these theoretical results are compared with
extensive numerical simulations. Finally, equilibrium fixed-point equations
are derived and compared with the thermodynamic approach based upon the
replica-symmetric mean-field approximation.
\end{abstract}
\newpage

\section{Introduction}
\pin
The parallel dynamics of extremely diluted asymmetric and layered
feedforward $Q\geq 2$-Ising neural networks have been solved exactly
(cfr.~\cite{DGZ}-\cite{BSV} and the references cited therein). This
has been possible because in these types of networks one knows that there
are no feedback loops as time progresses. In particular, this allows one
to derive recursion relations
for the relevant order parameters of these systems: the main
overlap for the condensed pattern, the mean of the neuron activities and
the variance of the residual overlap responsible for the intrinsic noise in
the dynamics of the main overlap (sometimes called the width-parameter).

These results are in strong contrast to those for the parallel
dynamics of networks
with symmetric connections. For these systems it turns out that even in the
diluted $Q=2$ case, feedback correlations become essential from the second
time step onwards, which already complicates the dynamics in a nontrivial
way \cite{WS}-\cite{PZ}.
For fully connected $Q=2$ systems, an increasing complexity of
such long-term temporal correlations makes the dynamics, in general,
extremely complicated.
Therefore, either approximate treatments of the feedback influence on the
network evolution or only the first few time steps of the main overlap
evolution have been analyzed so far. Nevertheless, this has led to some
important insights into the dynamics of the Little-Hopfield model
(cfr.~\cite{K}-\cite{OK} and references therein).

In this paper we consider the zero-temperature parallel dynamics of fully
connected $Q$-Ising neural networks for general $Q$. Generalizing a $Q=2$
result from the literature \cite{P}-\cite{VHK} we find that there exists a
Lyapunov function leading to the occurrence of fixed-points and two-cycles.
Since two-cycles in the $Q=2$ Little-Hopfield model seem to appear far from
the retrieval region \cite{FK} and/or seem to involve only a tiny fraction
of all spins \cite{SMK}, we only look at the fixed-point dynamics.

Using a probabilistic approach \cite{PZFC1} we extend our analysis for
extremely diluted \cite{BSVZ},\cite{BVZ} and layered $Q$-Ising networks
\cite{BSV}
to the non-trivial case of fully connected systems at zero temperature. In
particular, we develop a recursive scheme to calculate the relevant order
parameters of the system, i.e., the main overlap, the activity and the
variance of the residual overlap, for any time step. We write out these
expressions in detail for the first three time steps of the dynamics.
Furthermore, under the condition that the local field
becomes stationary we derive the fixed-point equations for these order
parameters. They are found to be the same as those derived via
thermodynamical methods \cite{BRS}. Finally, extensive numerical
simulations for the $Q=3$ model are compared with the theoretical results.

The rest of the paper is organized as follows. In Section \ref{sec:mod} we
introduce the model, its dynamics and the Hamming distance as a macroscopic
measure for the retrieval quality. In Section \ref{sec:gensch} we use the
probabilistic approach in order to derive a recursive scheme for the
evolution of the distribution of the local field, leading to recursion
relations for the order parameters of the model.
Using this general scheme, we explicitly calculate in Section
\ref{sec:ev.eq.} the order parameters for the first three time steps of
the dynamics. In Section \ref{sec:fixp} we show the existence of a
Lyapunov function at zero temperature and we discuss the evolution of
the system to fixed-point attractors.
A detailed discussion of the theoretical results obtained in
Section \ref{sec:ev.eq.} and a comparison with extensive numerical
simulations are presented in Section \ref{sec:results}. Some
concluding remarks are given in Section \ref{sec:con}.

\section{The model}
\label{sec:mod}
\pin
Consider a neural network $\Lambda$ consisting of $N$ neurons which can take
values $\sigma_i$ from a discrete set
        $ {\cal S} = \lbrace -1 = s_1 < s_2 < \ldots < s_Q
                = +1 \rbrace $.
Given the configuration
        ${\bsigma}_\Lambda(t)\equiv\{\sigma_j(t)\},
        j\in\Lambda=\{1,\ldots,N\}$,
the local field in neuron $i$ equals
\begin{equation}
        \label{eq:h}
        h_i({\bsigma}_{\Lambda\setminus \{i\}}(t))=
                \sum_{j\in\Lambda\setminus \{i\}} J_{ij}\sigma_j(t)
                         \,,
\end{equation}
with $J_{ij}$ the synaptic couplings between neurons $i$ and $j$.
In the sequel we write the shorthand notation $h_{\Lambda,i}(t) \equiv
h_i({\bsigma}_{\Lambda\setminus \{i\}}(t))$. The
configuration ${\bsigma}_\Lambda(t=0)$ is chosen as input. At
zero temperature all neurons are updated in parallel according to the rule
\begin{equation}
        \label{eq:enpot}
        \sigma_i(t)\rightarrow\sigma_i(t+1)=s_k:
                \min_{s\in{\cal S}} \epsilon_i[s|{\bsigma}_{\Lambda
                                  \setminus\{i\}}(t)]
            =\epsilon_i[s_k|{\bsigma}_{\Lambda \setminus\{i\}}(t)]
\,.
\end{equation}
Here the energy potential $\epsilon_i[s|{\bsigma}_{\Lambda\setminus\{i\}}]$
is defined by
\begin{equation}
        \epsilon_i[s|{\bsigma}_{\Lambda\setminus\{i\}}]=
                -\frac{1}{2}[h_i({\bsigma}_{\Lambda\setminus\{i\}})s-bs^2]
                                            \,,
\end{equation}
where $b>0$ is the gain parameter of the system. The updating rule
(\ref{eq:enpot}) is equivalent to using a gain function $\mbox{g}_b(\cdot)$,
\begin{eqnarray}
        \label{eq:gain}
        \sigma_i(t+1) &  =   &
               \mbox{g}_b(h_{\Lambda,i}(t))
                  \nonumber      \\
               \mbox{g}_b(x) &\equiv& \sum_{k=1}^Qs_k
                        \left[\theta\left[b(s_{k+1}+s_k)-x\right]-
                              \theta\left[b(s_k+s_{k-1})-x\right]
                        \right]
\end{eqnarray}
with $s_0\equiv -\infty$ and $s_{Q+1}\equiv +\infty$. For finite $Q$,
this gain function $\mbox{g}_b(\cdot)$ is a step function.
The gain parameter $b$ controls the average slope of $\mbox{g}_b(\cdot)$.

In this network we want to store $ p=\alpha N $ patterns. These patterns
are a collection of independent and identically distributed random
variables (i.i.d.r.v.), $\{{\xi}_i^\mu \in {\cal S}\}$,
$\mu \in {\cal P}=\{1,\ldots,p\}$ and   $i \in \Lambda$
with zero mean and variance $A=\Var[\xi_i^\mu]$. The synaptic couplings
between the neurons are chosen according to the Hebb learning rule
\begin{equation}
        \label{eq:J}
        J_{ij}=\frac{1}{NA} \sum_{\mu \in {\cal P}} \xi_i^\mu \xi_j^\mu
        \quad \mbox{for} \quad i \not=j, \quad J_{ii}=0           \,.
\end{equation}

To measure the retrieval quality of the system one can use the Hamming
distance between a stored pattern and the microscopic state of the network
\begin{equation}
        d({\bxi}^\mu,{\bsigma}_\Lambda(t))\equiv
                \frac{1}{N}
                \sum_{i\in\Lambda}[\xi_i^\mu-\sigma_i(t)]^2         \,.
\end{equation}
This naturally introduces the main overlap
\begin{equation}
        \label{eq:mdef}
        m_\Lambda^\mu(t)=\frac{1}{NA}
                \sum_{i\in\Lambda}\xi_i^\mu\sigma_i(t)
                \quad \mu \in {\cal P}
\end{equation}
and the arithmetic mean of the neuron activities
\begin{equation}
        \label{eq:adef}
        a_\Lambda(t)=\frac{1}{N}\sum_{i\in\Lambda}[\sigma_i(t)]^2    \,.
\end{equation}

\section{General dynamical scheme} \label{sec:gensch}
\pin
It is known that contrary to the asymmetrically diluted and layered
neural networks, the parallel dynamics of fully connected systems, even at
zero temperature, is not exactly solvable because of the strong feedback
correlations~\cite{BKS}.

On the basis of the probabilistic approach used before (see, e.g.,
\cite{BSV},\cite{PZFC1}) we develop in this section a
recursive dynamical scheme in order to calculate the distribution of
the local field at a general time step, for $Q \geq 2$-Ising neural
networks. This results in recursion relations determining the evolution of
the order parameters of these systems.

Suppose that the initial configuration of the network
$\{\sigma_i(0)\},{i\in\Lambda}$, is a collection of i.i.d.r.v.\ with mean
$\E[\sigma_i(0)]=0$, variance $\Var[\sigma_i(0)]=a_0$, and correlated with
only one stored pattern, say the first one $\{\xi^1_i\}$:
\begin{equation}
        \label{eq:init1}
        \E[\xi_i^\mu\sigma_i(0)]=\delta_{\mu,1}m^1_0 A
                \quad m^1_0>0 \, .
\end{equation}
This implies that by the law of large numbers (LLN) one gets for the main
overlap and the activity at $t=0$
\begin{eqnarray}
        m^1(0)&\equiv&\lim_{N \rightarrow \infty} m^1_\Lambda(0)
                \ustr{Pr}{=}\frac1A \E[\xi^1_i \sigma_i(0)]
                = m^1_0                                          \\
        a(0)&\equiv&\lim_{N \rightarrow \infty} a_\Lambda (0)
                \ustr{Pr}{=} \E[\sigma_i^2(0)]=a_0
\end{eqnarray}
where the convergence is in probability \cite{SH}. Using standard
signal-to-noise techniques (see, e.g., \cite{BVZ}), we find
the local field at $t=0$
\begin{eqnarray}
        \label{eq:init2}
        h_i(0)&=& \lim_{N \rightarrow \infty} \left[
                \xi_i^1 m_{\Lambda\setminus\{i\}}^1(0)
                + \frac1{NA}\sum_{\mu\in{\cal P}\setminus\{1\}}
                  \sum_{j\in\Lambda\setminus\{i\}}
                        \xi_i^\mu \xi_j^\mu\sigma_j(0) \right]
                        \nonumber                                     \\
         &\ustr{{\cal D}}{=}&\xi_i^1 m^1(0)+{\cal N}(0,\alpha a_0)  \,,
\end{eqnarray}
where the convergence is in distribution (see, e.g., \cite{SH}). The
quantity ${\cal N}(0,d)$ represents a Gaussian random variable with
mean $0$ and variance $d$.

For a general time step we find from eq.~(\ref{eq:gain}) and the
LLN in the limit $N \rightarrow \infty$ for the main overlap
(\ref{eq:mdef}) and the activity (\ref{eq:adef})
\begin{eqnarray}
        \label{eq:m}
        m^1(t+1) &\ustr{Pr}{=}& \frac{1}{A} \langle\!\langle
                 \xi_i^1\mbox{g}_b(h_i(t)) \rangle\!\rangle            \\
        \label{eq:a}
        a(t+1)   &\ustr{Pr}{=}& \langle\!\langle \mbox{g}_b^2(h_i(t))
                         \rangle\!\rangle
\end{eqnarray}
with $h_i(t) \equiv \lim_{N \rightarrow \infty} h_{\Lambda,i}(t)$.
In the above $\langle\!\langle \cdot \rangle\!\rangle$
denotes the average both over the distribution of the embedded patterns
$\{\xi_i^\mu\}$ and the initial configurations $\{\sigma_i(0)\}$. The
average over the initial configurations is hidden in an average over the
local field through the updating rule (\ref{eq:gain}). From the study of
layered networks(\cite{DKM}-\cite{BSV}) we know already that due to the
correlations there will be a third important parameter in the description
of the time evolution of the system: the influence of the non-condensed
patterns which is expressed by the variance of the residual
overlaps
\begin{eqnarray}
        D(t) &\equiv& \Var[r^\mu(t)] \, \\
        r^\mu(t) &\equiv& \lim_{N \rightarrow \infty}
        r_\Lambda^\mu(t) = \lim_{N \rightarrow \infty}
                \frac{1}{A\sqrt{N}}\sum_{i\in\Lambda}
                \xi_i^\mu\sigma_i(t)
                \quad \mu \in {\cal P}\setminus\{1\}                  \,.
        \label{eq:rdef}
\end{eqnarray}
Clearly, $D(0)= a_0/A$. The distribution of the embedded patterns is
given. It is the purpose of this section to calculate the distribution of
the local field as a function of time.

We start by rewriting the local field (\ref{eq:h}) at time $t$ in the
following way
\begin{eqnarray}
        \label{eq:hrecb}
        h_{\Lambda,i}(t)
            &=& \xi_i^1m_{\Lambda}^1(t) +
                  \frac{1}{NA}\sum_{j\in\Lambda}
                  \sum_{\mu \in {\cal P}}
                   \xi_i^\mu \xi_j^\mu \sigma_j(t)
                   -\alpha\sigma_i(t)                      \\
        \label{eq:hreca}
             &=& \xi_i^1m_{\Lambda}^1(t)-\alpha\sigma_i(t)+
                \frac{1}{\sqrt{N}}\sum_{\mu\in {\cal P}\setminus\{1\}}
                \xi_i^\mu r_{\Lambda}^\mu(t)                    \,.
\end{eqnarray}
{}From a technical point of view the explicit addition and subtraction of the
$\alpha \sigma_i(t)$ term in (\ref{eq:hrecb}) is convenient
in order to treat all indices in the sum over $j$ on an equal footing. This
turns out to be important to take into account all possible feedback loops.

We would like to remark that the set of $\alpha N$ variables $\{\xi_i^\mu
r_{\Lambda}^\mu(t)\}_\mu$ appearing in the last term of (\ref{eq:hreca})
are not independent because the $\xi_i^\mu$ are weakly dependent on the
$r_{\Lambda}^\nu(t), \, \nu \neq \mu$.
On the contrary, in the case of layered or diluted networks all terms of
this set of variables are independent such that their sum
is a normal distribution. Moreover, $r_{\Lambda\setminus\{i\}}^\mu(t)$ and
$\xi_i^\mu$ are also independent implying that the mean and the variance of
this distribution are known directly. The same is true for the fully
connected model at time $t=0$ where the network states $\{\sigma_i(0)\}_i$
are randomly distributed and independent of the non-condensed ($\mu>1$)
embedded patterns. But after applying the dynamics the $\sigma_i(t)$ and
the $\xi_i^\mu$ become dependent, leading to a weak dependence of
$r_{\Lambda \setminus \{i\}}^\mu(t)$ and $\xi_i^\mu$. This microscopic
dependence gives
rise to a macroscopic contribution after summing and taking the limit $N
\rightarrow \infty$. In this respect we mention that in ref.~\cite{AM} an
approximation has been put forward by neglecting precisely these
correlations between the $\xi_i^\mu$ and the $r_{\Lambda \setminus
\{i\}}^\mu(t)$. For an overview of improvements of this approximation for
$Q=2$ and corresponding numerical simulations we refer to \cite{OK}.

In order to determine the structure of the local field for fully connected
networks, we first concentrate on the evolution of the residual overlap
$r_\Lambda^\mu(t),\, \mu \in {\cal P}\setminus\{1\}$. As mentioned above the
dynamics induces the dependence of $\{\sigma_i(t)\}$ and $\{\xi_i^\mu\}$.
To study its consequences we rewrite the residual overlap (\ref{eq:rdef})
as
\begin{equation}
        r_\Lambda^\mu(t+1)=\frac{1}{\sqrt{N}A}\sum_{i \in \Lambda}
          \xi_i^\mu \mbox{g}_b(\hat h_{\Lambda,i}^\mu(t)+
             \frac{1}{\sqrt{N}}\xi_i^\mu r_\Lambda^\mu(t))
                \label{eq:f1}
\end{equation}
with
\begin{equation}
        \hat h_{\Lambda,i}^\mu(t)=h_{\Lambda,i}(t)-
                \frac{1}{\sqrt{N}}\xi_i^\mu r_\Lambda^\mu(t)            \,.
        \label{eq:f3}
\end{equation}
By subtracting the term $\xi_i^\mu r_\Lambda^\mu(t)/\sqrt{N}$ the modified
local field $\hat h_{\Lambda,i}^\mu(t)$ becomes only weakly dependent on
$\xi_i^\mu$, whereas $h_{\Lambda,i}(t)$ depends strongly on $\xi_i^\mu$.
The gain function $\mbox{g}_b(\cdot)$ is a step function that changes its
value by $s_{k+1}-s_{k}$ at $b(s_k+s_{k+1}), k=1,\ldots, Q-1$. Hence the
term $\xi_i^\mu r_\Lambda^\mu(t)/\sqrt{N}$ in (\ref{eq:f1}) becomes
relevant if for some $k$
\begin{equation}
        \label{eq:I_k}
        |\hat h_{\Lambda,i}^\mu(t)-b(s_k+s_{k+1})|<
                \frac{|\xi_i^\mu r_\Lambda^\mu(t)|}{\sqrt{N}}
                \quad k=1,\ldots, Q-1                             \,.
\end{equation}
Denoting by $I_k$ the set of indices satisfying condition (\ref{eq:I_k}),
we split the residual overlap into two sums:
\begin{eqnarray}
   && r_\Lambda^\mu(t+1)
       = \frac1{\sqrt{N}A}\sum_{i\notin\cup_kI_k}
                 \xi^\mu_i\mbox{g}_b(\hat h_{\Lambda,i}^\mu(t))
         + \frac1{\sqrt{N}A} \sum_{k=1}^{Q-1}\sum_{i\in I_k}\xi_i^\mu
                \left\{ \rule{0cm}{0.7cm}
                \mbox{g}_b(\hat h_{\Lambda,i}^\mu(t)) \right. \nonumber \\
     &&     \left. + \frac1{2}\left[
                \sign(b(s_k+s_{k+1})- \hat h_{\Lambda,i}^\mu(t))
                   +\sign(\frac1{\sqrt{N}}\xi_i^\mu r_\Lambda^\mu(t))\right]
                                (s_{k+1}-s_k)
                \right\}  \,. \nonumber \\
           \label{eq:f2}
\end{eqnarray}
In the argument of $\mbox{g}_b(\cdot)$ in the first term of (\ref{eq:f2})
the term $\xi_i^\mu r_\Lambda^\mu(t)/\sqrt{N}$ is left out since it
can not change the value of $\mbox{g}_b(\cdot)$ by definition of the sets
$I_k$.
Combining the first two terms, eq.~(\ref{eq:f2}) can be rewritten as
\begin{eqnarray}
    && r_\Lambda^\mu(t+1)=
        \frac1{\sqrt{N}A}\sum_{i\in\Lambda} \xi_i^\mu
                \mbox{g}_b(\hat h_{\Lambda,i}^\mu(t))
     +\frac1{\sqrt{N}A} \sum_{k=1}^{Q-1}\sum_{i\in I_k}
            \xi_i^\mu
                   \nonumber        \\
    && \times        \frac1{2}
             \left[\sign(b(s_k+s_{k+1})- \hat h_{\Lambda,i}^\mu(t))+
                \sign(\frac1{\sqrt{N}}\xi_i^\mu r_\Lambda^\mu(t) )
                        \right] (s_{k+1}-s_k)  \, . \nonumber \\
           \label{eq:rreca}
\end{eqnarray}
We then consider the limit $N \rightarrow \infty$. In this limit the
cardinal number of the set $I_k$ becomes deterministic
\begin{equation}
        \lim_{N \rightarrow \infty}\frac{|I_k|}{\sqrt{N}}=
                2|\xi_i^\mu r^\mu(t)|
                f_{\hat h_i^\mu (t)}(b(s_{k+1}+s_k))
\end{equation}
with $f_{\hat h_i^\mu (t)}$ the probability density of the modified local
field $\hat h_i^\mu (t)$ at time $t$. We remark that in the thermodynamic
limit the density distribution of the modified local field $\hat
h_i^\mu(t)$  at time $t$ equals the density distribution of the local field
$h_i(t)$ itself.

Furthermore, we apply the CLT on the first term of (\ref{eq:rreca}) and the
LLN on the second term with the random variable $r_\Lambda^\mu(t)$ fixed.
This yields the following result
\begin{eqnarray}
        r^\mu(t+1) = \tilde r^\mu(t)+\chi(t)r^\mu(t)
           \label{eq:rrec}
\end{eqnarray}
where, recalling eqs.~(\ref{eq:a}) and (\ref{eq:f3})
\begin{equation}
        \tilde r^\mu(t) \equiv \lim_{N \rightarrow \infty}
           \frac1{A\sqrt{N}}\sum_{i\in \Lambda} \xi_i^\mu
                \mbox{g}_b(\hat h_{\Lambda , i}^\mu(t))
                \ustr{{\cal D}}{=} {\cal N}_\mu(0,a(t+1)/A)
           \label{eq:w}
\end{equation}
because of the weak dependence of $\hat h_{i}^\mu(t)$
and $\xi_i^\mu$, and
\begin{equation}
        \chi(t) = \sum_{k=1}^{Q-1} f_{\hat h_i^\mu (t)}(b(s_{k+1}+s_k))
                   (s_{k+1}-s_k) \,.
            \label{eq:chi}
\end{equation}

{}From the relation (\ref{eq:rrec}) one finds a recursion relation for the
variance of the residual overlap
\begin{equation}
        \label{eq:Drec}
        D(t+1)=\frac{a(t+1)}{A}+\chi^2(t)D(t)+
                2\chi(t)\Cov[\tilde r^\mu(t),r^\mu(t)] \,.
 \label{eq:f4}
\end{equation}
At this point it is interesting to remark that the last term on the r.h.s.\
of (\ref{eq:f4}) is entirely coming from the correlations caused by the
fully connected structure of the network. It is absent for layered (compare
eq.~(30c) of ref.~\cite{BSV}) and hence, of course, also for extremely
diluted asymmetric architectures. In the latter case also the second term
on the r.h.s. of (\ref{eq:f4}) disappears.

Starting from the local field at time $t+1$ in the form of
eq.~(\ref{eq:hreca}) and using expressions (\ref{eq:rrec}), (\ref{eq:w})
and (\ref{eq:chi}) we obtain in the limit $N \rightarrow \infty$, after
some straightforward manipulations
\begin{equation}
        h_i(t+1)=\xi_i^1m^1(t+1)+
                \chi(t) [h_i(t)-\xi_i^1m^1(t)+\alpha\sigma_i(t)]
                + {\cal N}(0,\alpha a(t+1)) \, .
        \label{eq:hrec}
\end{equation}
{}From this it is clear that the local field at time $t+1$ consists out of a
discrete part and a normally distributed part, viz.
\begin{equation}
        h_i(t)=M_i(t) + {\cal N}(0,V(t))
\end{equation}
where $M_i(t)$ satisfies the recursion relation
\begin{equation}
        M_i(t+1)=\chi(t) [M_i(t)-\xi_i^1m^1(t)+\alpha\sigma_i(t)]
                         + \xi_i^1m^1(t+1)
     \label{eq:Mrec}
\end{equation}
and
\begin{equation}
        V(t+1)=\alpha A D(t+1)
     \label{eq:Vrec}
\end{equation}
with $D(t+1)$ given by the recursion relation (\ref{eq:f4}).

We still have to determine $f_{h_i(t)}$ in eq.~(\ref{eq:chi}). We know that
the quantity $M_i(t)$ consists out of the signal term and a discrete noise
term, viz. \begin{equation}
        M_i(t)=\xi _i^1 m^1(t) +  \sum_{t'=0}^{t-1} \alpha
         \left[\prod_{s=t'}^{t-1} \chi(s)\right] \, \sigma _i(t')  \,.
\end{equation}
The evolution equation tells us that $\sigma _i(t')$ can be replaced by
$g_b(h_i(t'-1))$ such that the second term of $M_i(t)$ is the sum of
stepfunctions of correlated variables. These are also correlated through
the dynamics with the normally distributed
part of $h_i(t)$. Therefore the local field can be considered as a
transformation of a set of correlated normally distributed variables
$x_s,\, s=0,\ldots,t-2,t$. Defining the correlation matrix $C_{s s'}\equiv
\E[x_s x_{s'}]$ we arrive at the following expression for the
probability density of the local field at time $t$
\begin{eqnarray}
     f_{h_i(t)}(y)&=&\int\prod_{s=0}^{t-2} dx_s dx_t ~
             \delta \left(y - M_i(t)-\sqrt{V(t)}\,x_t\right) \nonumber\\
             &\times& \frac{1}{\sqrt{\mbox{det}(2\pi C)V(t)}}
            ~\mbox{exp}\left(-\frac{1}{2}{\bf x}^T C^{-1}
            {\bf x}\right)
            \label{eq:fhdis}
\end{eqnarray}
with ${\bf x}=(x_0,\ldots x_{t-2},x_t)$.

Together with the eqs.~(\ref{eq:m})-(\ref{eq:a}) for $m^1(t+1)$ and
$a(t+1)$ the equations (\ref{eq:rrec})-(\ref{eq:f4}), (\ref{eq:Mrec}) and
(\ref{eq:fhdis}) form a recursive scheme in order to obtain the
order parameters of the system. The practical difficulty which remains is
the explicit calculation of the correlations in the network at different
time steps as present in eq. (\ref{eq:f4}).

\section{Evolution equations up to the third time step}
        \label{sec:ev.eq.}
\pin
Following the general recursive scheme established in Section
\ref{sec:gensch} evolution equations are derived for the order parameters
of a fully connected $Q$-Ising network for the first three time steps,
taking into account all correlations. This generalizes and extends the
$Q=2$ results in the literature mentioned in the Introduction.

\subsection{First step dynamics} \label{sec:1st}
\pin
Starting from eqs.~(\ref{eq:init2}),(\ref{eq:m}) and (\ref{eq:a}) one has
immediately
\begin{eqnarray}
        m^1(1)&=& \frac1{A}\left\langle\!\left\langle
            \xi^1 \int {\cal D}z ~ \mbox{g}_b(\xi^1m^1(0)
              +\sqrt{\alpha A D(0)}\,z) \right\rangle\!\right\rangle
                           \\
        a(1)  &=& \left\langle\!\left\langle\int
           {\cal D}z ~ \mbox{g}_b^2(\xi^1m^1(0)
              +\sqrt{\alpha A D(0)}\,z) \right\rangle\!\right\rangle  \,,
\end{eqnarray}
where $\left\langle\!\left\langle \cdots \right\rangle\!\right\rangle$ now
stands for the average taken with respect to the distribution of the
first pattern and the initial configuration and ${\cal D}z$ denotes a
Gaussian measure ${\cal D}z=dz
\exp(-\frac1{2}z^2)/\sqrt{2\pi}$. We recall that $D(0)=a_0/A$. Next, from
the initial conditions (\ref{eq:init1})-(\ref{eq:init2})
and the definition of the modified local field (\ref{eq:f3}) one also knows
that $\{\xi_i^\mu, \mbox{g}_b(\hat h_i^\mu(0))\}_i$ and $\{\xi_i^\mu,
\sigma_i(0)\}_i$
become a set of uncorrelated parameters for $\mu \in {\cal P} \setminus
\{1\}$. Here $\hat h_i^\mu(t) \equiv \lim_{N \rightarrow \infty} \hat
h^\mu_{\Lambda,i}(t)$. Therefore
\begin{equation}
        \label{eq:cor w^mu r^mu}
        \Cov[{\tilde r}^\mu(0), r^\mu(0)] =
        \E[\sigma_i(0) \mbox{g}_b(\hat h_i^\mu(0))]            \,.
\end{equation}
Using the recursion relation (\ref{eq:Drec}) this leads to
\begin{equation}
        D(1)=\frac{a(1)}{A} + \chi^2(0) D(0) +
                2 \frac{\chi(0)}{A}
                   \left\langle\!\left\langle
                   \sigma(0)\int {\cal D} z ~
                    \mbox{g}_b(\xi^1m^1(0)+\sqrt{\alpha A D(0)}\,z)
                        \right\rangle\!\right\rangle
\end{equation}
with
\begin{equation}
        \chi(0)= \left\langle\!\left\langle
              \frac1{\sqrt{\alpha A D(0)}} \int {\cal D} z ~
               z ~ \mbox{g}_b(\xi^1m^1(0)+\sqrt{\alpha A D(0)}\,z)
                 \right\rangle\!\right\rangle                       \,.
\end{equation}
These results generalize the corresponding $Q=2$ results (see, e.g.,
\cite{K} and \cite{PZFC1}).

\subsection{Second step dynamics} \label{sec:2nd}
\pin
First we need the distribution of the local field at time $t=1$. This
follows immediately from eqs.~(\ref{eq:Mrec}) and (\ref{eq:Vrec})
\begin{eqnarray}
        \label{eq:h(1)}
        h_i(1) &=& \xi_i^1m^1(1)+\chi(0)
                [h_i(0)-\xi_i^1m^1(0)+\alpha\sigma_i(0)]
                + {\cal N}(0,\alpha a(1))                         \,.
\end{eqnarray}
Recalling again eqs.~(\ref{eq:m}) and (\ref{eq:a}), the main overlap and
the activity read
\begin{eqnarray}
        m^1(2)&=& \frac1{A}
          \left\langle\!\left\langle
               \xi^1\int {\cal D} z ~ \mbox{g}_b \left(\xi^1m^1(1)+
              \alpha\chi(0)\sigma(0)+\sqrt{\alpha A D(1)}\,z \right)
                \right\rangle\!\right\rangle    \\
        a(2)  &=& \left\langle\!\left\langle
               \int {\cal D} z ~ \mbox{g}_b^2\left(\xi^1m^1(0)+
               \alpha\chi(0)\sigma(0)+\sqrt{\alpha A D(1)}\,z \right)
                  \right\rangle\!\right\rangle
\end{eqnarray}
and
\begin{equation}
        \chi(1)=\frac1{\sqrt{\alpha A D(1)}}
                \left\langle\!\left\langle
                \int {\cal D} z ~ z ~
                \mbox{g}_b\left(\xi^1m^1(1)+\alpha\chi(0)\sigma(0)+
                        \sqrt{\alpha A D(1)}\,z \right)
                \right\rangle\!\right\rangle   \,.
\end{equation}

These equations correspond to the equations for the $Q=2$-network found
in \cite{GDM}. The calculation of the third order parameter, i.e., the
variance of the residual overlap, needs some more work. From the
recursion formula (\ref{eq:rrec}) one finds
\begin{eqnarray}
        \label{eq:D2a}
        \Cov[{\tilde r}^\mu(1),r^\mu(1)]
           &=& \Cov[{\tilde r}^\mu(1), {\tilde r}^\mu(0)]
                       +\chi(0)\Cov[{\tilde r}^\mu(1),r^\mu(0)]
\\ &=& \frac1A R(2,1) + \chi(0) \frac1A R(2,0)
\end{eqnarray}
with the correlation parameters, $R(t,\tilde t)$, defined as
\begin{eqnarray}
        R(t,\tilde t) &=& \langle\!\langle
                           \mbox{g}_b(\hat h^\mu(t-1)) ~
                           \mbox{g}_b(\hat h^\mu(\tilde t-1))
                        \rangle\!\rangle \quad t,\tilde t \geq 1
                \label{eq:Rt_ttilde}       \\
        R(t,0) &=& \langle\!\langle
                   \sigma(0)  \mbox{g}_b(\hat h^\mu(t-1))
                  \rangle\!\rangle \quad t \geq 1         \,.
        \label{eq:R(t,0)}
\end{eqnarray}
This is based on the fact that by definition of the modified local field
(\ref{eq:f3}) $\{\xi_i^\mu, \mbox{g}_b(\hat h_i^\mu(0))\}_i$ and
$\{\xi_i^\mu, \sigma_i(0)\}_i$ become a set of uncorrelated variables.
These results lead to the recursion relation (recall eq.~(\ref{eq:Drec}))
\begin{equation}
        \label{eq:D2}
        D(2)=\frac{a(2)}{A} +\chi^2(1)D(1)
            +2\frac{\chi(1)}{A}\left(R(2,1)+\chi(0)R(2,0)
                         \right) \,.
\end{equation}
We still have to determine the $R(t,\tilde t)$. The correlation $R(2,0)$
can be written down immediately again by using the definition of the
modified local field at $t=1$
\begin{equation}
        R(2,0)=\left\langle\!\left\langle
                 \sigma(0) \int {\cal D} z ~
                  \mbox{g}_b\left(\xi^1m^1(1)+\alpha\chi(0)\sigma(0)+
                        \sqrt{\alpha A D(1)}\,z \right)
                \right\rangle\!\right\rangle      \,.
                \end{equation}
To obtain $R(2,1)$, one remarks that due to the dependence of
$\sigma_i(0)$ and $\sigma_i(1)$ the local fields $h_i(1)$ and
$h_i(0)$ are correlated. The correlation coefficient of their normally
distributed part in general defined as
\begin{equation}
  \rho(t,\tilde t) \equiv \frac{\E[(h(t)-M(t))(h(\tilde t)-M(\tilde t))]}
                            {\sqrt{V(t)} \sqrt{V(\tilde t)}}
\end{equation}
is found using the recursion formula (\ref{eq:h(1)})
\begin{equation}
        \rho(1,0)=\frac{\alpha R(1,0) + \alpha A \chi(0) D(0)}
                         {\sqrt{\alpha A D(0)}\sqrt{\alpha A D(1)}}     \,.
             \label{eq:for10}
\end{equation}
Employing all this in eq.~(\ref{eq:Rt_ttilde}) we arrive at
\begin{eqnarray}
     R(2,1)&=&\int {\cal D}w^{1,0}(x,y) \,\,
                  \mbox{g}_b\left(\xi^1m^1(0)+\sqrt{\alpha A D(0)}\,x
                            \right)  \nonumber \\
           &\times& \mbox{g}_b\left(\xi^1m^1(1)+\alpha\chi(0)\sigma(0)+
                                \sqrt{\alpha A D(1)}\,y
                             \right)   \, .
\end{eqnarray}
Here the joint distribution ${\cal D}w^{1,0}(x,y)$ equals
\begin{eqnarray}
        {\cal D}w^{1,0}(x,y)&=&\frac{dx~dy}{2\pi\sqrt{1-\rho(1,0)^2}}
                \exp\left(-\frac{x^2-2\rho(1,0)xy+y^2}
                                {2(1-\rho(1,0)^2)}
                    \right) \, .
\end{eqnarray}

We remark that, for $Q=2$, the result (\ref{eq:D2}) is slightly different
from the corresponding result in \cite{PZFC2} (see, e.g., their eq.(39)).
In more detail, in their approach these authors make an ansatz stating the
independence of the normally distributed and discrete part in the noise
arising from $t \geq 2 $ onwards. They explicitly state that they have no
convincing arguments in favour of (as well as against) this ansatz for
$t \geq 2$. In our approach we do not need this ansatz.
This results in a more complicated expression for $R(2,1)$ than the
corresponding one found in \cite{PZFC2}, indicating that this ansatz really
ignores some correlations. In fact both expressions coincide if we put
$\rho(1,0)$ equal to zero.

\subsection{Third step dynamics} \label{sec:3d}
\pin
We start by writing down the distribution of the local field at time $t=2$.
{}From eqs.~(\ref{eq:Mrec}) and (\ref{eq:Vrec}) we find
\begin{equation}
        \label{eq:h(2)}
        h_i(2)=  \xi_i^1m^1(2)+\alpha\chi(1)[\sigma_i(1)+
             \chi(0)\sigma_i(0)] +{\cal N}(0,\alpha A D(2)) \,.
\end{equation}
This gives for the main overlap
\begin{equation}
        \label{eq:m(3)a}
        m^1(3)=\frac{1}{A}\left\langle\!\left\langle
                \xi^1 \mbox{g}_b\left(\xi^1m^1(2)+\alpha\chi(1)
                \left[\sigma(1)+\chi(0)\sigma(0)\right]
                        +\sqrt{\alpha A D(2)}\,y \right)
                \right\rangle\!\right\rangle
\end{equation}
with $y$ the Gaussian random variable ${\cal N}(0,1)$. The average has to
be  taken over $y,~\sigma_i(0)$ and $\sigma_i(1)$. The average over
$\sigma_i(0)$ causes no difficulties because this initial configuration
is chosen randomly. The average over $y$, the Gaussian random variable
appearing in $h_i(2)$, and $\sigma_i(1)$ is more tricky because
$h_i(2)$ and $\sigma_i(1)$ are correlated by the dynamics.
However, the evolution equation (\ref{eq:gain}) tells us that $\sigma_i(1)$
can be replaced by $\mbox{g}_b(h_i(0))$ and, hence, the average taken over
$h_i(0)$ instead of $\sigma_i(1)$.

{}From the recursion relation (\ref{eq:hrec}) one finds for the correlation
coefficient between $h_i(0)$ and $h_i(2)$
\begin{equation}
     \rho(2,0)=\frac{\alpha \left[
                      R(2,0)+\chi(1)\left[R(1,0)+\chi(0)a(0) \right]
                 \right]}
                {\sqrt{\alpha A D(0)}\sqrt{\alpha A D(2)}} \, .
                \label{eq:for20}
\end{equation}
Using all this the main overlap at the third time step (\ref{eq:m(3)a})
becomes
\begin{eqnarray}
        \label{eq:m(3)}
        m^1(3)&=& \frac{1}{A}
                \left\langle\!\left\langle
                \xi^1\int {\cal D}w^{2,0}(x,y)\,\mbox{g}_b
                \left(\xi^1m^1(2) +  \right.\right.\right.
                    \nonumber \\
         && \left.\left.\left. \alpha\chi(1)
                  \left[\mbox{g}_b(\xi^1m^1(0)+\sqrt{\alpha A D(0)}\,x)
                                +\chi(0)\sigma(0)
                        \right]
               +\sqrt{\alpha A D(2)}\,y
                \right)
                \right\rangle\!\right\rangle  \nonumber \\
\end{eqnarray}
where the joint distribution of $x$ and $y$ equals
\begin{equation}
        \label{eq:DH20}
        {\cal D}w^{2,0}(x,y)=
                \frac{dx~dy}{2\pi\sqrt{1-\rho(2,0)^2}}
                \exp\left(-\frac{x^2-2\rho(2,0)xy+y^2}
                                {2(1-\rho(2,0)^2)}
                    \right) \,.
\end{equation}
In an analogous way one arrives at the expression for the activity  at the
third time step
\begin{eqnarray}
        \label{eq:a3}
        a(3)&=&\left\langle\!\left\langle
           \int {\cal D}w^{2,0}(x,y)\,\mbox{g}_b^2
                \left(\xi^1m^1(2) +  \right.\right.\right.
               \nonumber \\
             && \left.\left.\left. \alpha\chi(1)
                   \left[\mbox{g}_b(\xi^1m^1(0)+\sqrt{\alpha A D(0)}\,x)
                                +\chi(0)\sigma(0)
                        \right]
                        +\sqrt{\alpha A D(2)}\,y
                \right)
              \right\rangle\!\right\rangle  \, . \nonumber \\
\end{eqnarray}
In order to find the variance of the residual overlap at the third
time step, $D(3)$, we start by rewriting eq.~(\ref{eq:Drec}) as
\begin{equation}
        \label{eq:D3}
        D(3)=\frac{a(3)}{A} + \chi^2(2)
                +2 \frac{\chi(2)}{A}
                        \left(R(3,2)+ \chi(1)
                        \left(R(3,1)+\chi(0)R(3,0)\right)
                        \right)
\end{equation}
with
\begin{eqnarray}
        \chi (2) &=& \langle\!\langle
           \frac{1}{\sqrt{\alpha A D(2) (1-\rho (2,0)^2)}}
           \int {\cal D}z\,z\,\int{\cal D}y\,
           g_b\left(\xi ^1 m(2) \right.  \nonumber\\
           &+&
           \alpha \chi (1)
           \left(g_b(\xi ^1 m(0) + \sqrt{\alpha A D(0)}\,y)
           + \chi (0)\sigma (0)\right) \nonumber\\
           &+& \sqrt{\alpha A D(2) (1-\rho (2,0)^2)}\,z
           + \sqrt{\alpha A D(2)}\, \rho (2,0)y \left. \right)
        \rangle\!\rangle \, .
\end{eqnarray}
Here we have used the recursion relation (\ref{eq:rrec}) for $r^\mu(2)$
and the fact that $\{\xi_i^\mu, \mbox{g}_b(\hat h_i^\mu(0))\}_i$ and
$\{\xi_i^\mu, \sigma_i(0)\}_i$ become a collection of uncorrelated
variables for $\mu \in {\cal P}\setminus\{1\}$.
We then have to calculate the correlations $R(3,0)$, $R(3,1)$ and $R(3,2)$.
{}From the definition (\ref{eq:Rt_ttilde}), the local field (\ref{eq:h(2)})
and the joint distribution (\ref{eq:DH20}) one easily arrives at
\begin{eqnarray}
        R(3,0)&=&\left\langle\!\left\langle
                \sigma(0) \int {\cal D} w^{2,0}(x,y)\,
                \mbox{g}_b\left(\xi^1m^1(2) + \right. \right. \right.
                \nonumber\\
              &&\hspace{-1.8cm} \left. \left.\left. \alpha\chi(1)
                        \left[g_b(\xi^1m^1(0)+\sqrt{\alpha A D(0)}\,x)
                                +\chi(0)\sigma(0)
                        \right]
                        +\sqrt{\alpha A D(2)}\,y
                \right)
               \right\rangle\!\right\rangle      \\
      R(3,1)&=&\left\langle\!\left\langle
                \int {\cal D} w^{2,0}(x,y)\,
                \mbox{g}_b\left(\xi^1m^1(0) + \sqrt{\alpha A D(0)}\,x
                  \right) \mbox{g}_b\left(\xi^1m^1(2) +
                \right. \right. \right.  \nonumber\\
              &&\hspace{-1.8cm} \left.\left.\left. \alpha\chi(1)
                        \left[\mbox{g}_b(\xi^1m^1(0)+
                        \sqrt{\alpha A D(0)}\,x) +\chi(0)\sigma(0)
                        \right]
                        +\sqrt{\alpha A D(2)}\,y
                \right)
               \right\rangle\!\right\rangle \,.
\end{eqnarray}
Finding $R(3,2)$ is more tricky since, after rewriting the network
configurations $\{\sigma_i(t)\}$ at time $t=1$ and $t=2$ by means of the
gain function (\ref{eq:gain}), the local fields at the three first time
steps appear. So one has to calculate the elements of the correlation
matrix of these local fields in general defined by
\begin{equation}
    C(t,\tilde t) \equiv \sqrt{\alpha A D(t)}\sqrt{\alpha A D(\tilde t)}
                              \,  \rho(t,\tilde t)   \,.
\end{equation}
The correlation coefficients $\rho(1,0)$ and $\rho(2,0)$ have been
calculated already before (recall eqs.~(\ref{eq:for10})
and (\ref{eq:for20})). The correlation coefficient $\rho(2,1)$ of $h(2)$
and $h(1)$ is found by using the recursion relation (\ref{eq:hrec}) as
\begin{equation}
        \rho(2,1)=\frac{\alpha
            \left[ R(2,1)+\chi(0)R(2,0) +\chi(1) A D(1)\right]}
                {\sqrt{\alpha A D(2)}\sqrt{\alpha A D(1)}}      \,.
\end{equation}
The distribution function ${\cal D}w^{2,1}(x,y,z)$ of the three local
fields equals
\begin{equation}
         {\cal D}w^{2,1}(x,y,z)=
                \frac{dx~dy~dz}{(2\pi)^{3/2}\sqrt{\mbox{Det} w^{2,1}}}
                \exp\left(-\frac12
                    \left(\begin{array}{l}x\\y\\z\end{array}\right)
                        (w^{2,1})^{-1}
                    \left(\begin{array}{rcl}x&\!y&\!z\end{array}\right)
                    \right)
\end{equation}
where
\begin{equation}
       w^{2,1}=
   {\left(
        \begin{array}{lcr}
\!\!\alpha A D(0)\!\!   & C(1,0)        \!\!    & C(2,0)        \!\!\\
\!\!C(1,0)      \!\!    & \alpha A D(1) \!\!    & C(2,1)        \!\!\\
\!\!C(2,0)      \!\!    & C(2,1)        \!\!    & \alpha A D(2) \!\!
        \end{array}
   \right)}                                                              \,.
\end{equation}
Finally, using all this information one gets for the correlation parameter
\begin{eqnarray}
        \label{eq:R(3,2)}
        R(3,2)&=&\left\langle\!\left\langle\int {\cal D}w^{2,1}(x,y,z)\,
                \mbox{g}_b\left(\xi^1m^1(1)+\alpha\chi(0)\sigma(0)+y
                    \right)  \right.\right.
                    \nonumber      \\
              && \hspace{-1cm}\left.\left. \mbox{g}_b\left(\xi^1m^1(2)+
                        \alpha\chi(1)
                     \left[\mbox{g}_b\left(\xi^1m^1(0)+x\right)
                                     +\chi(0)\sigma(0)
                        \right]
                        +z
                    \right)
              \right\rangle\!\right\rangle \,.
\end{eqnarray}
These results can be compared with those for extremely diluted systems.
If the dilution is symmetric (see refs.~\cite{WS},\cite{PZ} for the
case $Q=2$) feedback loops over two time steps can exist, but the
probability to have loops over a longer time period equals zero. Therefore
the $\sigma_i(0)$-term in (\ref{eq:h(2)}) drops out.
Furthermore in the $Q=2$ case the expression for the correlation
coefficient (\ref{eq:for20}) simply reads $\rho(2,0)=R(2,0)$.
If the dilution is asymmetric \cite{BSVZ}, all feedback disappears and the
local field is simply Gaussian distributed.

\section{Fixed-point equations} \label{sec:fixp}
\pin
A second type of results can be obtained by requiring through the recursion
relations (\ref{eq:Drec}), (\ref{eq:Mrec}) and (\ref{eq:Vrec}) that the
local field becomes stationary. We show that this leads to the same
fixed-point equations as those found from thermodynamics in \cite{BRS}.

For the Q-Ising model at zero temperature one can show that
\begin{equation}
        H(t)=-\frac14\sum_{i\in\Lambda}
                \left(\sum_{j\in\Lambda} J_{ij}\sigma_j(t)\tilde\sigma_i(t)
                -b(\tilde\sigma_i^2(t)+\sigma_i^2(t))
                           \right)
\end{equation}
with $\tilde\sigma_i(t)$ chosen such that
\begin{equation}
        \epsilon_i[\tilde\sigma_i(t)|{\bsigma}_{\Lambda \setminus\{i\}}(t)]
         = \min_{s\in{\cal S}} \epsilon_i[s|{\bsigma}_{\Lambda
                                  \setminus\{i\}}(t)]\, ,
\end{equation}
is a Lyapunov function. For finite $N$, $H(t)$ is bounded from below
implying
that $H(t+1)-H(t)=0$ after finitely many time steps. This can be realized
for $\sigma_i(t+2)=\sigma_i(t) ~ \forall i\in\Lambda$.
The proof is straightforward and completely analogous to the argumentation
used in \cite{P},\cite{VHK}.
Both a fixed point and a two-cycle satisfy this condition. As stated in the
introduction we only study fixed-points.

Since the evolution equations for the order parameters in the extremely
diluted and layered $Q$-Ising models
do not change their form as time progresses, the fixed-point equations are
obtained immediately by leaving out the time dependence
(see \cite{BSVZ},\cite{BSV}). This still allows small fluctuations in the
configurations $\{\sigma_i\}$.

Since in the fully connected model treated here the form of the evolution
equations  for the order parameters do change by the explicit appearance of
the $\{\sigma_i(t)\}, t \geq 0 $, we can not use that procedure to obtain
the fixed-point equations. Instead we require that the distribution of
the local field becomes independent of time. This is a stronger condition
because fluctuations in the network configuration are no longer allowed.
Consequently, the main overlap and activity in the fixed-point are found
from the definitions (\ref{eq:mdef}), (\ref{eq:adef}) and not from leaving
out the time dependence in the recursion relation (\ref{eq:m}) and
(\ref{eq:a}). The same line of reasoning is followed in, e.g.,
\cite{SFa},\cite{SFb}.

We start by eliminating the time-dependence in the evolution equations for
the local field (\ref{eq:hrec}). This leads to
\begin{equation}
        \label{eq:hfix}
        h_i=\xi_i^1m^1+\frac1{1-\chi}\dN(0,\alpha a)
                +\alpha\frac{\chi}{1-\chi}\sigma_i
\end{equation}
with $h_i \equiv \lim_{t \rightarrow \infty} h_i(t)$.
This expression consists out of two parts: a normally distributed part
$\tilde h_i = {\cal N}(\xi_i^1m^1,\alpha a / (1-\chi)^2)$
and some discrete noise part. We remark that this
discrete noise coming from the correlation of the $\{\sigma_i(t)\}$ at
different time steps is inherent in the fully connected dynamics.

Employing the expression eq.~(\ref{eq:hfix}) in the updating rule
(\ref{eq:gain}) one finds
\begin{equation}
        \label{eq:sfp}
        \sigma_i=\mbox{g}_b(\tilde h_i+\alpha\eta\sigma_i) \quad
                \eta=\chi/(1-\chi)                                      \,.
\end{equation}
This is a self-consistent equation in $\sigma_i$ which
in general admits more than one solution. This type of equation has been
solved in the case of analog neural networks with continous time dynamics
using a Maxwell construction \cite{SFa},\cite{SFb}.
Such a construction is standard in thermodynamics in order
to maximize the exponent of the integrand appearing in free energy
calculations.
Here we use a similar geometrical construction to treat eq.~(\ref{eq:sfp}).

Let $L$ be the straight line which connects the centers of the plateaus
of the gain function $\mbox{g}_b(\cdot)$.
The equations for the functions $\mbox{g}_b(\cdot)$ and $L(\cdot)$ read
\begin{eqnarray}
        \label{eq:g}
        \mbox{g}_b&:&x \mapsto s_k
                ~\mbox {if}~  b(s_k+s_{k-1})<x<b(s_k+s_{k+1})   \\
        L  &:&x \mapsto \frac{x}{2b}                             \,.
\end{eqnarray}
The condition on the r.h.s. of (\ref{eq:g}) is a condition on
$x$. Using the definition of $L(\cdot)$, one can transform this into
a condition on the image of $L(\cdot)$,
${\cal I}_L=\{y\in\Rbar\:|\: \exists \:x\in\Rbar:L(x)=y\}$, viz.
\begin{equation}
        \label{eq:gL}
        \mbox{g}_b(x)=s_k ~
                \mbox{if}~
               \frac{s_k+s_{k-1}}{2}<L(x)<\frac{s_k+s_{k+1}}{2}    \,.
\end{equation}
Consider the transformation ${\cal T}:(x,y)\mapsto (x-\alpha\eta y,y)$
\begin{equation}
        {\cal T}(L):x \mapsto \frac{x}{2(b-\frac{\alpha\eta}{2})}    \,.
\end{equation}

The function ${\cal T}(\mbox{g}_b)(\cdot)$ is not bijective while
${\cal T}$ is not one-to-one. To obtain a unique solution for
eq.~(\ref{eq:sfp}) we modify the former function such that it becomes a
step function with the
same step height as the one in ${\cal T}(\mbox{g}_b)(\cdot)$ and the width
of the steps such that ${\cal T}(L)$ connects the centers of the plateaus:
\begin{equation}
        {\cal T}_L(\mbox{g}_b)(x)=s_k
                ~\mbox{if}~
                        (b-\frac{\alpha\eta}{2})(s_k+s_{k-1})
                        <x<
                        (b-\frac{\alpha\eta}{2})(s_k+s_{k+1})
\end{equation}
or, using (\ref{eq:gain})
\begin{equation}
        \label{eq:Tg}
        {\cal T}_L(\mbox{g}_b)(x)=\mbox{g}_{\tilde b}(x)
                ~\mbox{with}~
                \tilde b = b-\frac{\alpha\eta}{2}                       \,.
\end{equation}
This at first sight ad-hoc modification leads us to a unique solution
of the self-consistent equation (\ref{eq:sfp}).
Indeed, from this modified transformation we know that
\begin{equation}
        \mbox{g}_b(\tilde h + \alpha\eta\sigma)
                \simeq
        \mbox{g}_{\tilde b}(\tilde h + \alpha\eta\sigma
                -\alpha\eta \mbox{g}_b(\tilde h + \alpha\eta\sigma))
                                                \,,
\end{equation}
such that
\begin{equation}
        \sigma_i=\mbox{g}_{\tilde b}(\tilde h_i)
\,.
\end{equation}

At this point we remark that plugging this result into the local field
equation (\ref{eq:hfix}) tells us that the latter is the sum of two
Gaussians with shifted mean (see also \cite{HO}).

Using the definition of the main overlap and activity
(\ref{eq:mdef}) and (\ref{eq:adef}) in the limit $N \rightarrow \infty$,
one finds in the fixed point
\begin{eqnarray}
        \label{eq:m1fix}
        m^1 &=&\left\langle\!\left\langle\xi^1\int {\cal D}
        z ~  \mbox{g}_{\tilde b}
                \left( \xi^1m^1 + \sqrt{\alpha A D}\,z
                \right)\right\rangle\!\right\rangle
          \\
        \label{eq:afix}
        a   &=&\left\langle\!\left\langle\int {\cal D}
        z ~  \mbox{g}_{\tilde b}^2
                \left( \xi^1m^1 + \sqrt{\alpha A D}\,z
                \right)\right\rangle\!\right\rangle
          \,.\end{eqnarray}
{}From (\ref{eq:rrec}), (\ref{eq:Drec}) and (\ref{eq:chi}) it is clear that
\begin{equation}
        \label{eq:Dfix}
        D =\frac{a/A}{(1-\chi)^2}
\end{equation}
with
\begin{equation}
        \label{eq:chifix}
        \chi=\frac1{\sqrt{\alpha A D}}
                \left\langle\!\left\langle\int {\cal D}
                z ~  z \, \mbox{g}_{\tilde b}
                        \left( \xi^1m^1 + \sqrt{\alpha A D}\,z
                        \right)\right\rangle\!\right\rangle \,.
\end{equation}
These resulting equations (\ref{eq:m1fix})-(\ref{eq:Dfix}) are the same as
the fixed-point equations derived from a
replica-symmetric mean-field theory treatment in \cite{BRS}. Their solution
leads to the $\alpha-b$ phase diagram Fig.~1b in \cite{BRS}.
We end with the observation that for analog networks
the construction (\ref{eq:g})-(\ref{eq:Tg}) is not necessary: the
fixed-point equation (\ref{eq:sfp}) has only one solution.

\section{Numerical simulations} \label{sec:results}
\pin
As an illustrative example the equations derived in Section \ref{sec:ev.eq.}
have been worked out explicitly in the case of the $Q=3$ model with
equidistant states and a uniform distribution of the patterns ($A=2/3$).

For this model a thermodynamic replica-symmetric mean-field theory approach
leads to a capacity-gain phase diagram discussed already in \cite{BRS}
(Fig.~1b). As explained in Section \ref{sec:fixp} the same phase diagram
can be obtained through the dynamical approach presented here. For
convenience and completeness this phase diagram is reproduced here as
Fig.~1. At this point it is also useful to recall that there are two types
of retrieval states. In region I the mean-square random overlap with the
non-condensed patterns, $r$, is of order $O(1)$ while in region II $r$ is
of order $O(10)$ \cite{BRS}.

For specific network parameters corresponding to different points in the
retrieval region of this equilibrium phase diagram,
indicated as $1$ to $4$, we have compared the dynamics governed by the
evolution equations found here with extensive simulations involving
system-sizes up to $N=6000$ (each data point is averaged over 1600 runs).

Figures~2-5 present an overview of these results by
plotting the overlap $m^1(t)$, the activity $a(t)$ and the Hamming distance
$d(t)$ versus the initial overlap $m^1_0$ with the condensed pattern.
(We forget about the superscript 1). The initial activity is taken to be
$a_0=0.85$.

First we consider region I.
For network parameters corresponding to point 1 below the thermodynamic
transition line, i.e., $\alpha=0.005, b=0.3$, we see in Fig.~2 that for
$m_0 \geq 0.33$ the dynamics quickly evolves to an overlap $m=1$ and that
the Hamming distance is zero for $m_0 \geq 0.37$. The activity attains the
value $2/3$, meaning that the network configuration is uniformly distributed.
The boundary between the $m=1$ attractor and the zero-attractor is rather
sharply determined.

For a network corresponding to point 2 above the thermodynamic transition
line, with $\alpha=0.03, b=0.5$, we need a larger value of $m_0$ to reach
the $m=1$ attractor and a Hamming distance zero. As seen in Fig.~3, $m_0$
has to be at least $0.75$. Also the boundary between the $m=1$ attractor
and the zero-attractor is less sharply determined. Figure~4 shows that this
behavior is qualitatively the same for $\alpha=0.009, b=0.7$, corresponding
to point 3 situated above the spin-glass transition in the phase diagram.
In this case the value of $m_0$ has to be at least $0.85$. For the other
network parameters we have looked at, e.g., $\alpha=0.0115, b=0.5$ the
global behavior is similar.

For network parameters corresponding to points in region II of the phase
diagram, e.g., point 4 with $\alpha=0.015, b=0.1$ it is shown in Fig.~5
that the main overlap goes to its maximum value for almost all values of
$m_0$. The basin of attraction of the zero fixed-point is zero.
The activity, however, goes to a value larger than $2/3$.
The network configuration is no longer uniformly distributed: the state
$\sigma_i=0$ has a smaller probability to appear than the states
$\sigma_i=\pm1$. Hence, the Hamming distance is never zero. This must be
due to the fact that the influence of the non-condensed patterns is much
larger here ($r \approx O(10)$). The same qualitative behavior is found
for network parameters corresponding to points in region II below the
thermodynamic transition line, e.g., $\alpha= 0.005, b=0.1$.

\section{Concluding remarks} \label{sec:con}
\pin
In this paper we have derived the evolution equation for the distribution
of the local field governing the parallel dynamics at zero temperature
of fully connected $Q$-Ising networks, taking into account {\it all}
feedback correlations.
This leads to a general recursive scheme which allows us to calculate the
relevant order parameters of the system, i.e., the main overlap, the
activity and the variance of the residual overlap, for any time step.
We have worked out this scheme explicitly for the first three time steps of
the dynamics.

Under the condition that the local field becomes stationary
we have also obtained the fixed-point equations for these order parameters.
They are found to be the same as those derived via thermodynamic methods
\cite{BRS}.

As an illustration we have presented a detailed discussion of these
results for the $Q=3$-model and we have made a comparison with extensive
numerical simulations. It is seen that these numerical results provide
excellent support for our theoretical predictions and that the first three
time steps do give already a clear picture of the time evolution in the
retrieval regime of the network.

\section*{Acknowledgments}
\pin
This work has been supported in part by the Research Fund of the
K.U.Leuven (Grant OT/94/9) and the Korea Science and Engineering
Foundation through the SRC program. The authors are indebted to S.~Amari,
R.~K\"uhn A.~Patrick and V.~Zagrebnov for constructive discussions.
One of us (D.B.) thanks the Belgian National Fund for Scientific
Research for financial support.



\section*{Figures}
\begin{figure}[h]
\caption{The $\alpha-b$ phase diagram (see [21] figure 1.b).}
\end{figure}

\vspace{-5.cm}

\begin{figure}[h]
\epsfxsize=5.cm
\centerline{\rotate[r]{\epsfbox[50 50 600 750]{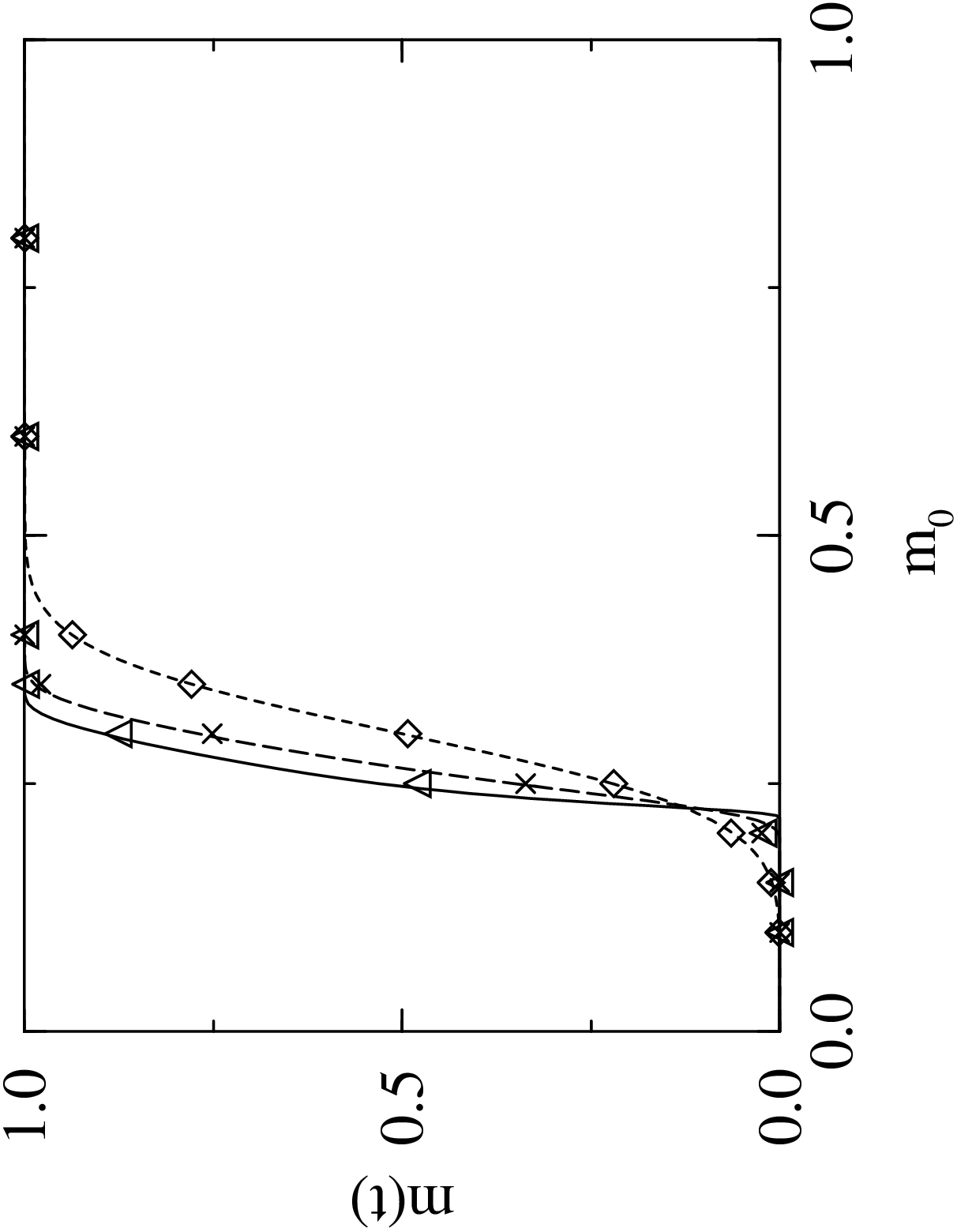}}}
\centerline{\rotate[r]{\epsfbox[50 50 600 750]{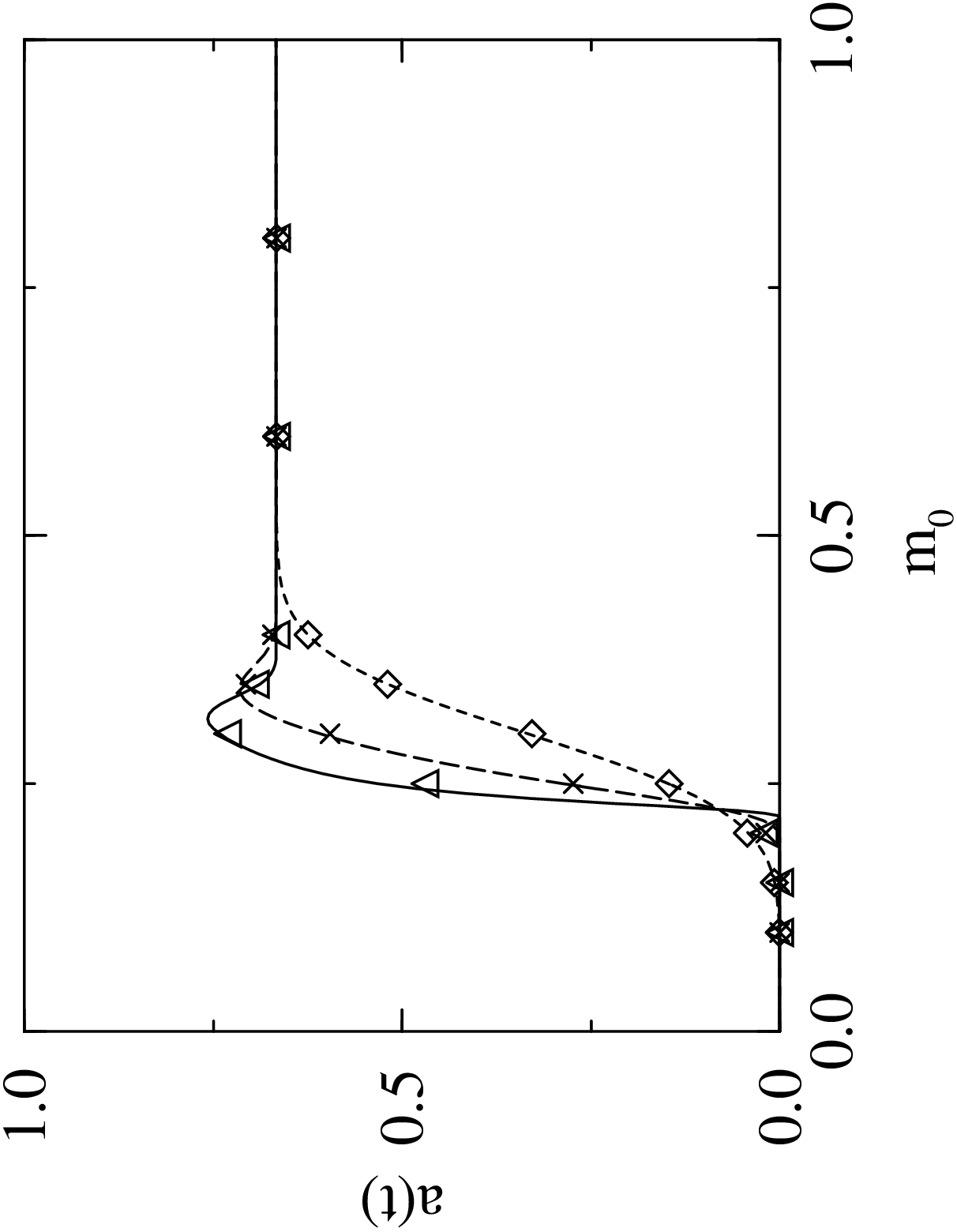}}}
\centerline{\rotate[r]{\epsfbox[50 50 600 750]{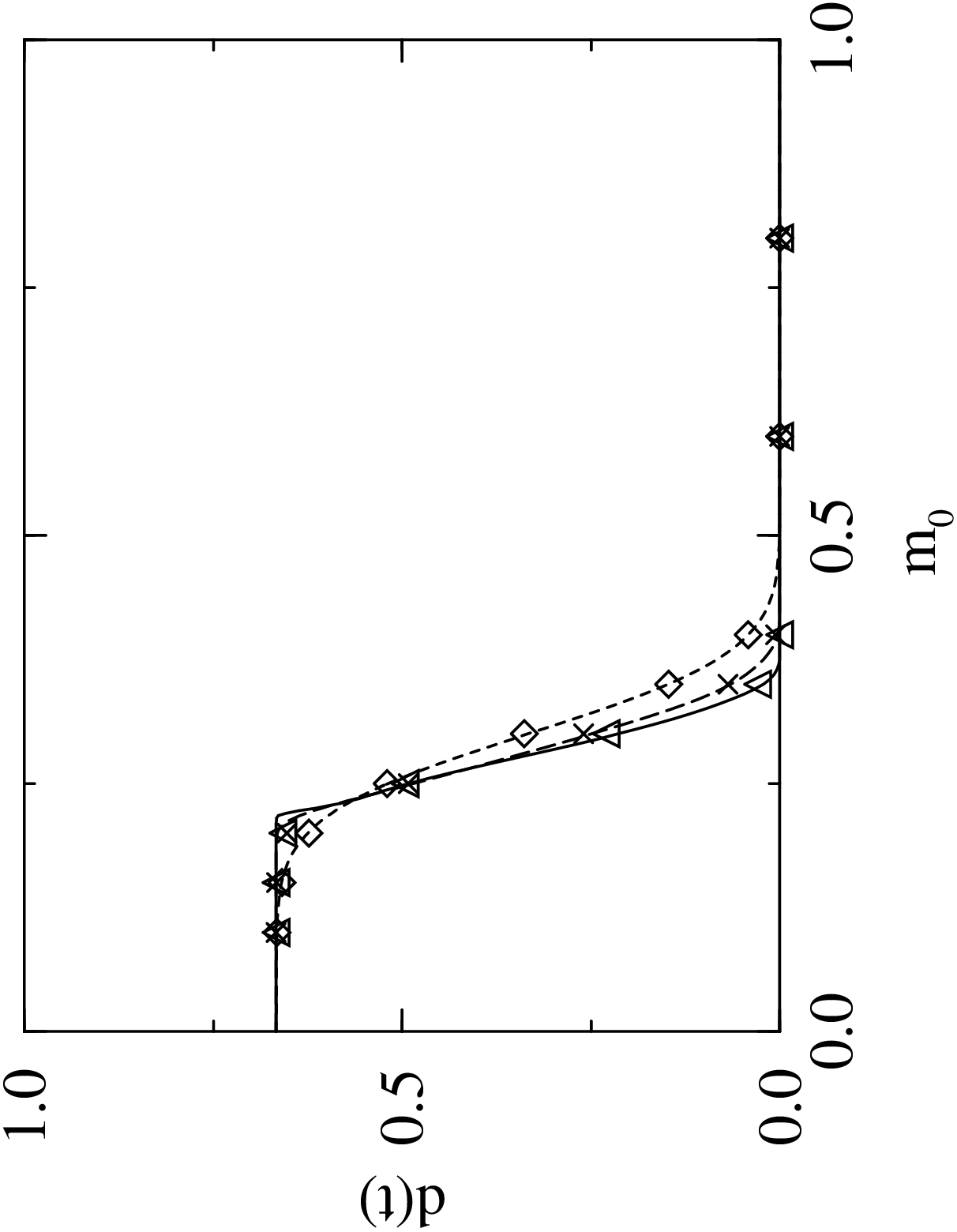}}}
\caption{
A comparison of the theoretical results and numerical simulations
for systems with $N=6000$. The overlap $m(t)$, the activity $a(t)$ and
the Hamming distance $d(t)$ are presented for the first three time steps
as a function of $m_0$ for the network parameters $b=0.3, \alpha= 0.005,
a_0=0.85$.
Theoretical (simulations) results for the first, second and third time step
are indicated by a short-dashed curve (diamond symbol), a long-dashed curve
(times symbol) and a full line (triangle symbol) respectively.}
\end{figure}

\begin{figure}[t]
\epsfxsize=5.cm
\centerline{\rotate[r]{\epsfbox[50 50 600 750]{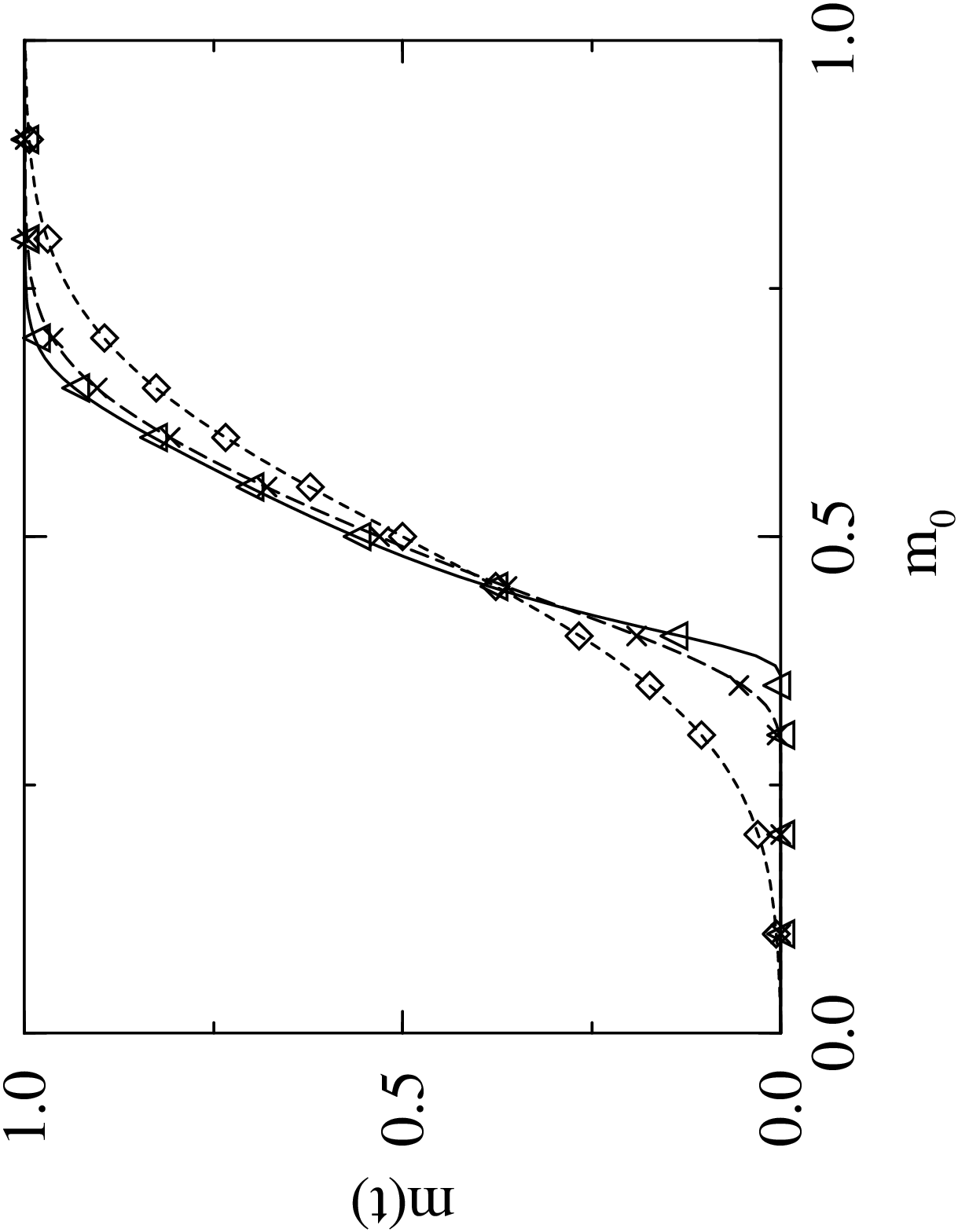}}}
\centerline{\rotate[r]{\epsfbox[50 50 600 750]{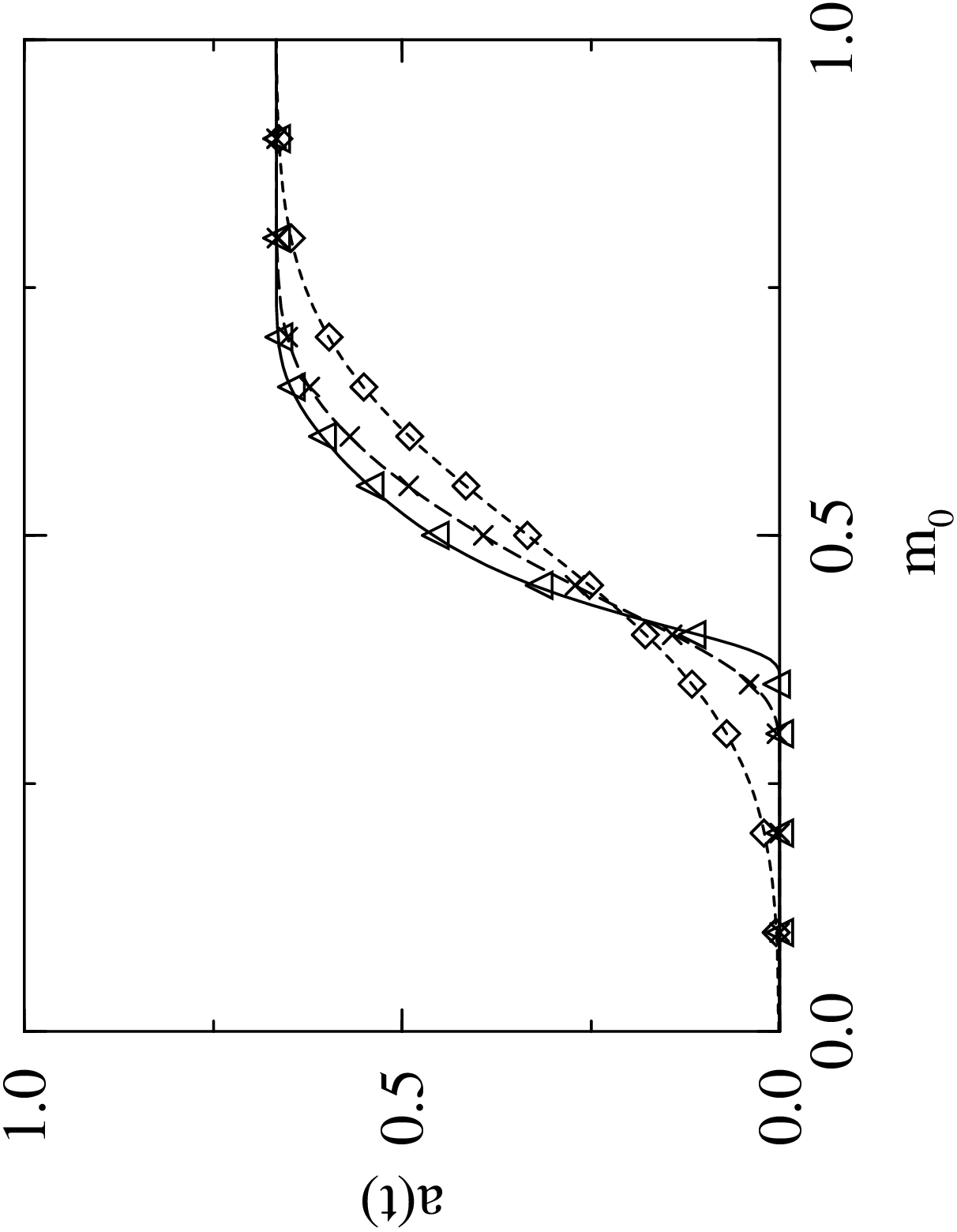}}}
\centerline{\rotate[r]{\epsfbox[50 50 600 750]{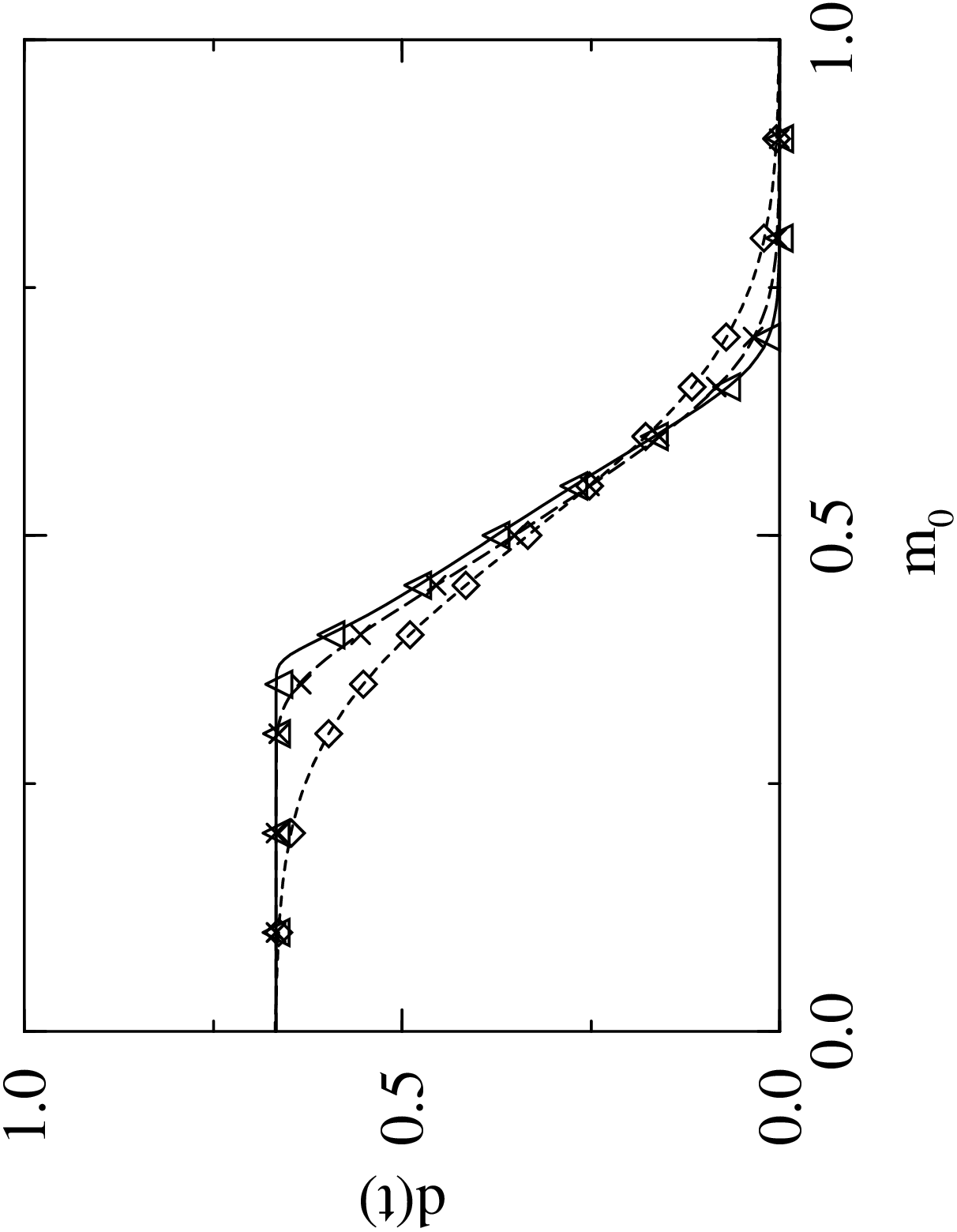}}}
\caption{
As in Fig. 2, for the network parameters $b=0.5, \alpha= 0.03, a_0=0.85$.}
\end{figure}

\begin{figure}[t]
\epsfxsize=5.cm
\centerline{\rotate[r]{\epsfbox[50 50 600 750]{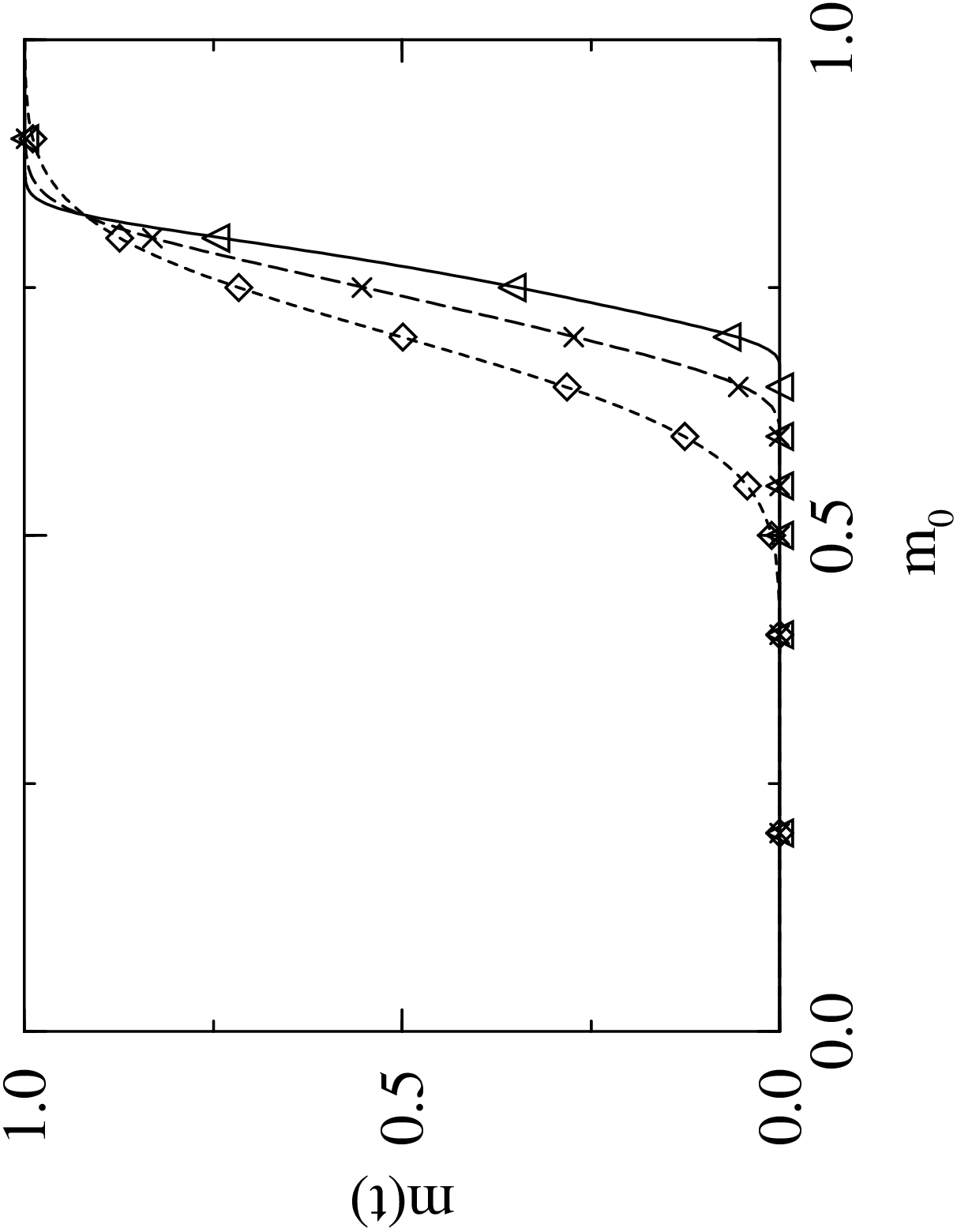}}}
\centerline{\rotate[r]{\epsfbox[50 50 600 750]{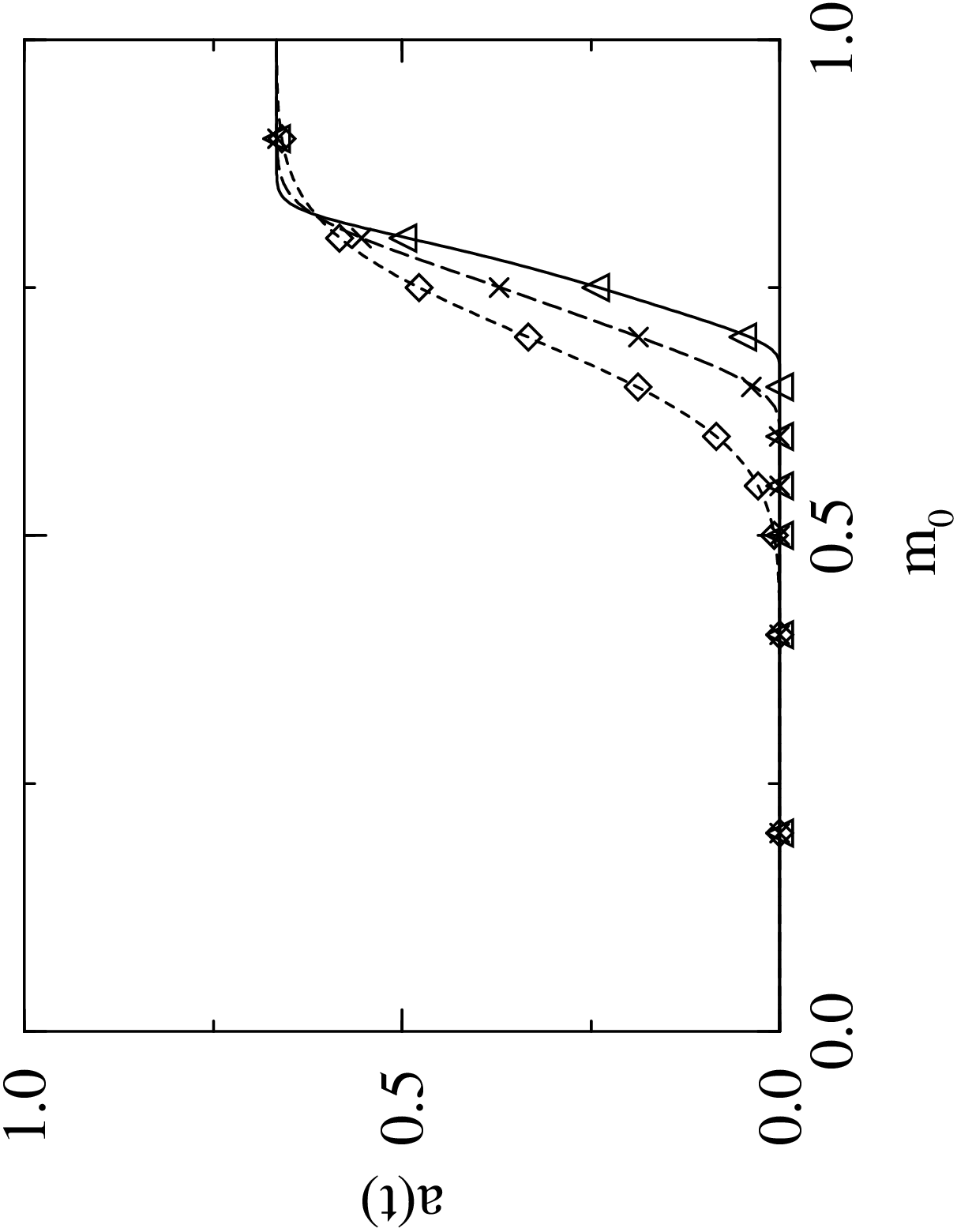}}}
\centerline{\rotate[r]{\epsfbox[50 50 600 750]{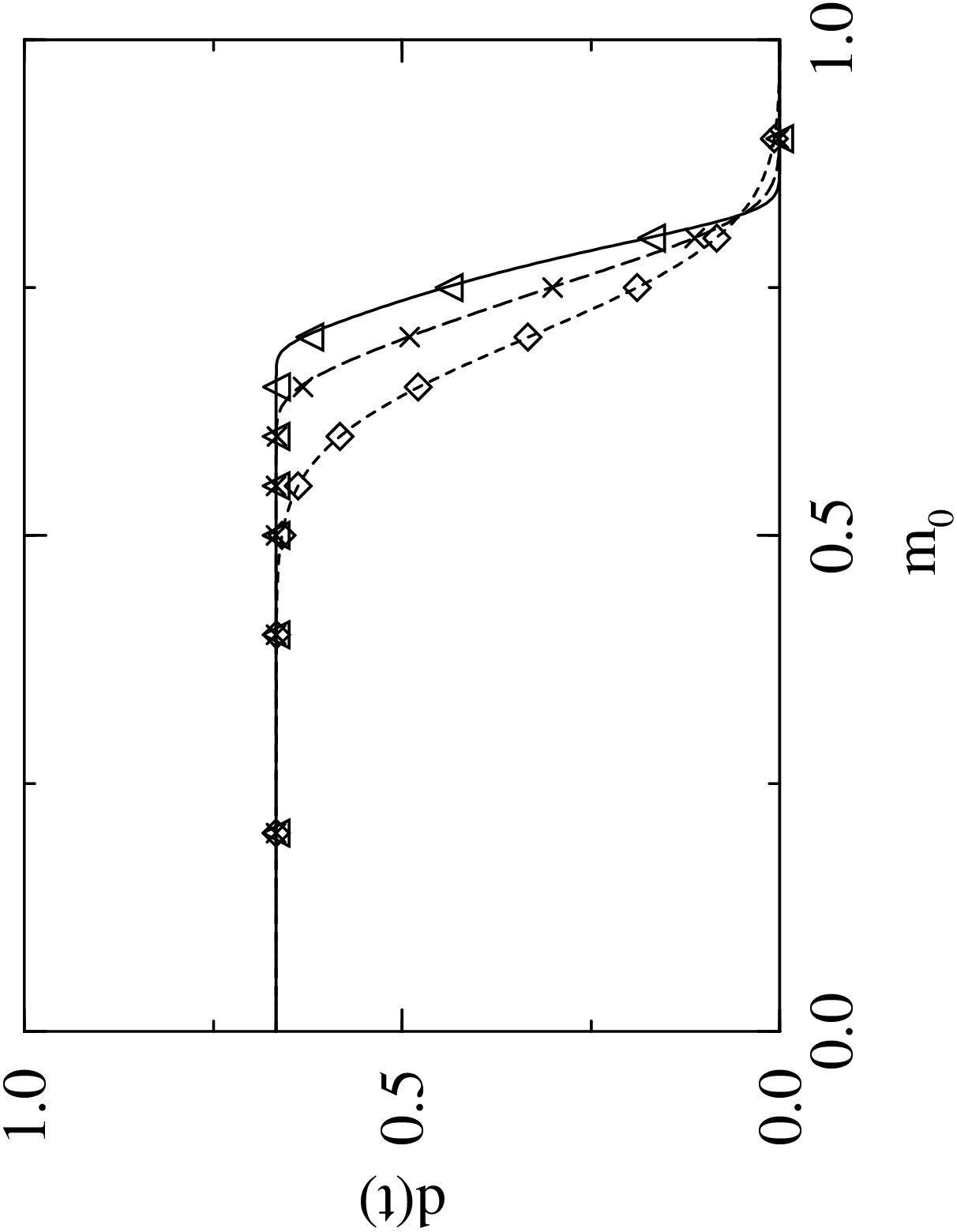}}}
\caption{
As in Fig. 2, for the network parameters $b=0.7, \alpha= 0.009, a_0=0.85$.}
\end{figure}

\begin{figure}[t]
\epsfxsize=5.cm
\centerline{\rotate[r]{\epsfbox[50 50 600 750]{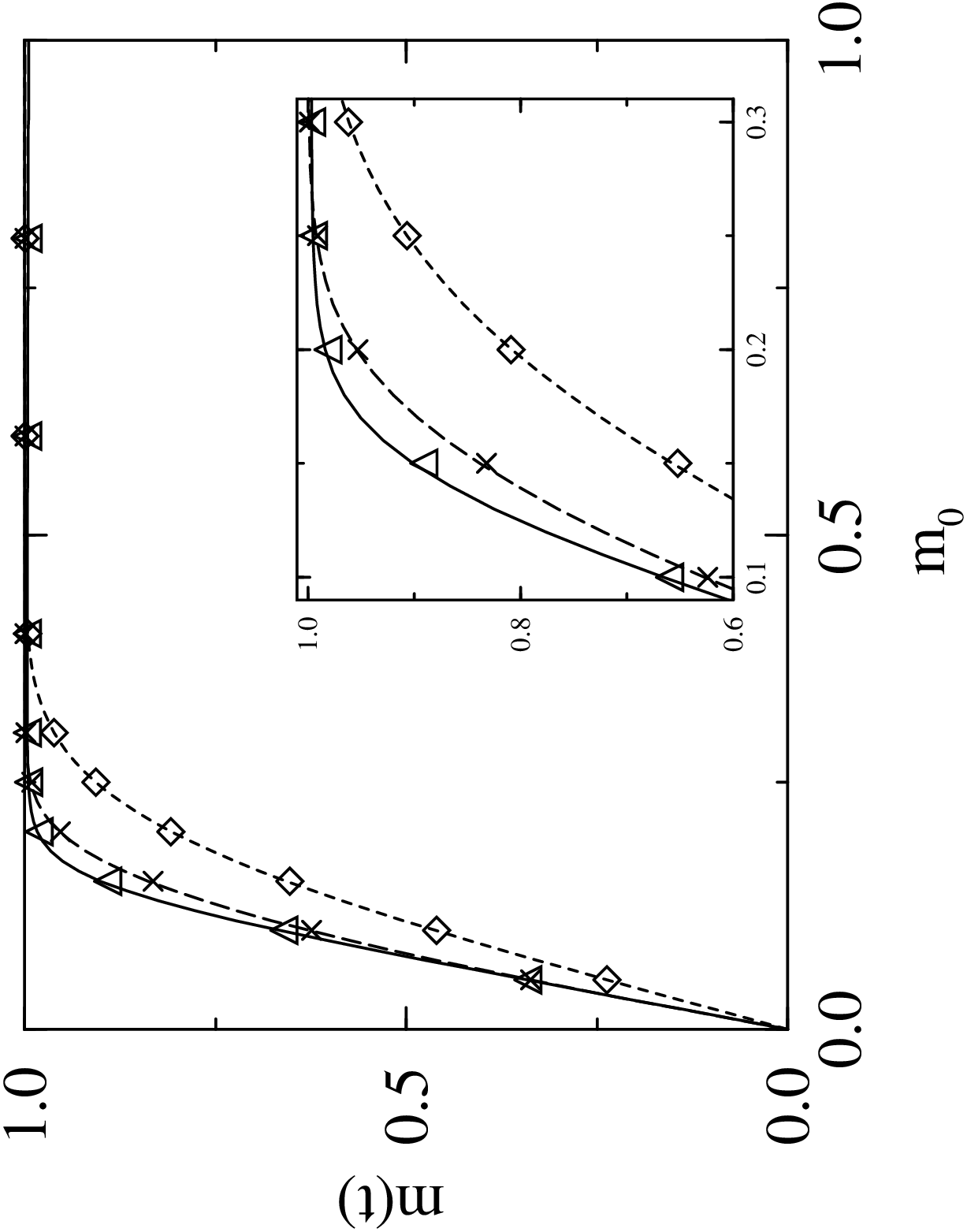}}}
\centerline{\rotate[r]{\epsfbox[50 50 600 750]{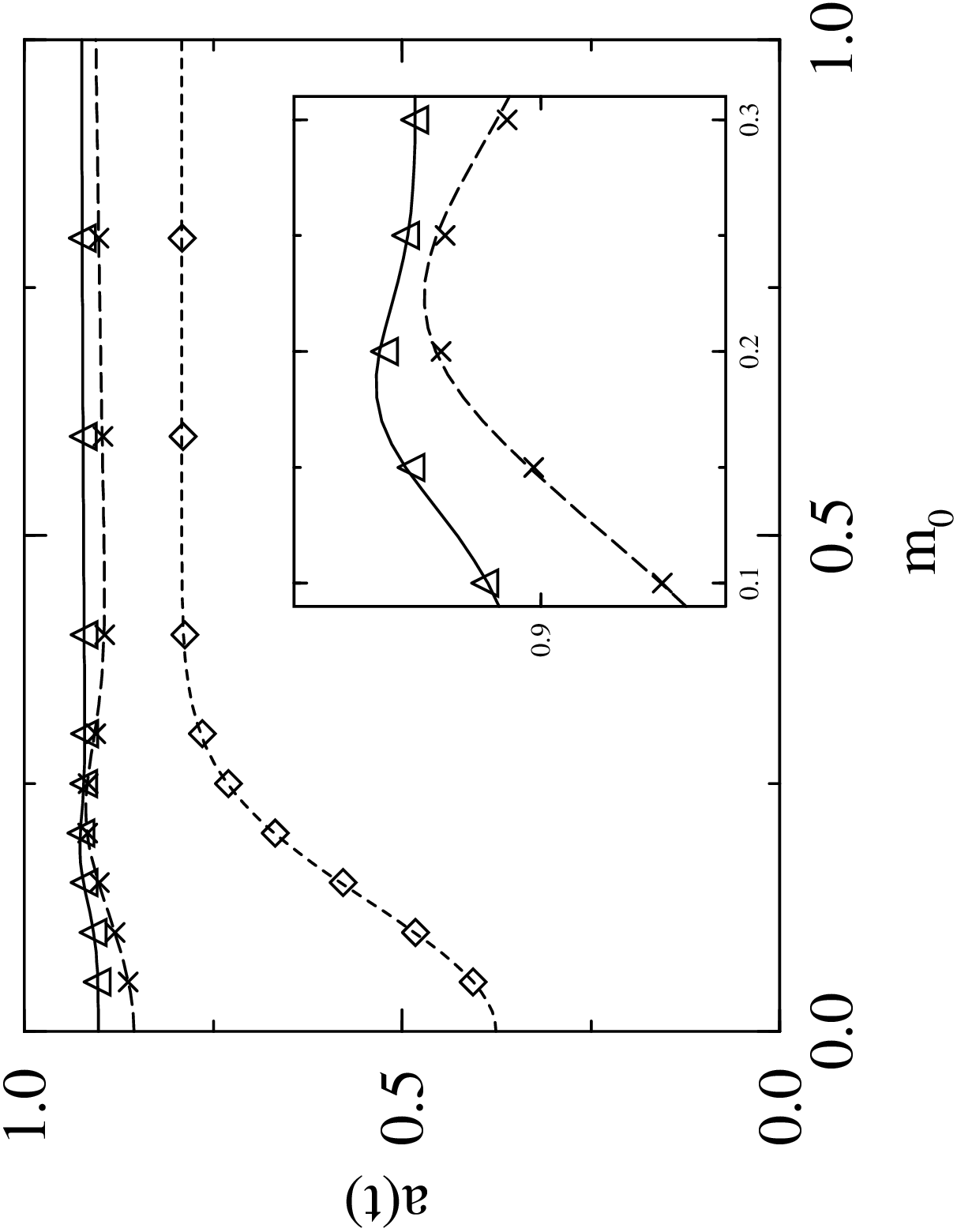}}}
\centerline{\rotate[r]{\epsfbox[50 50 600 750]{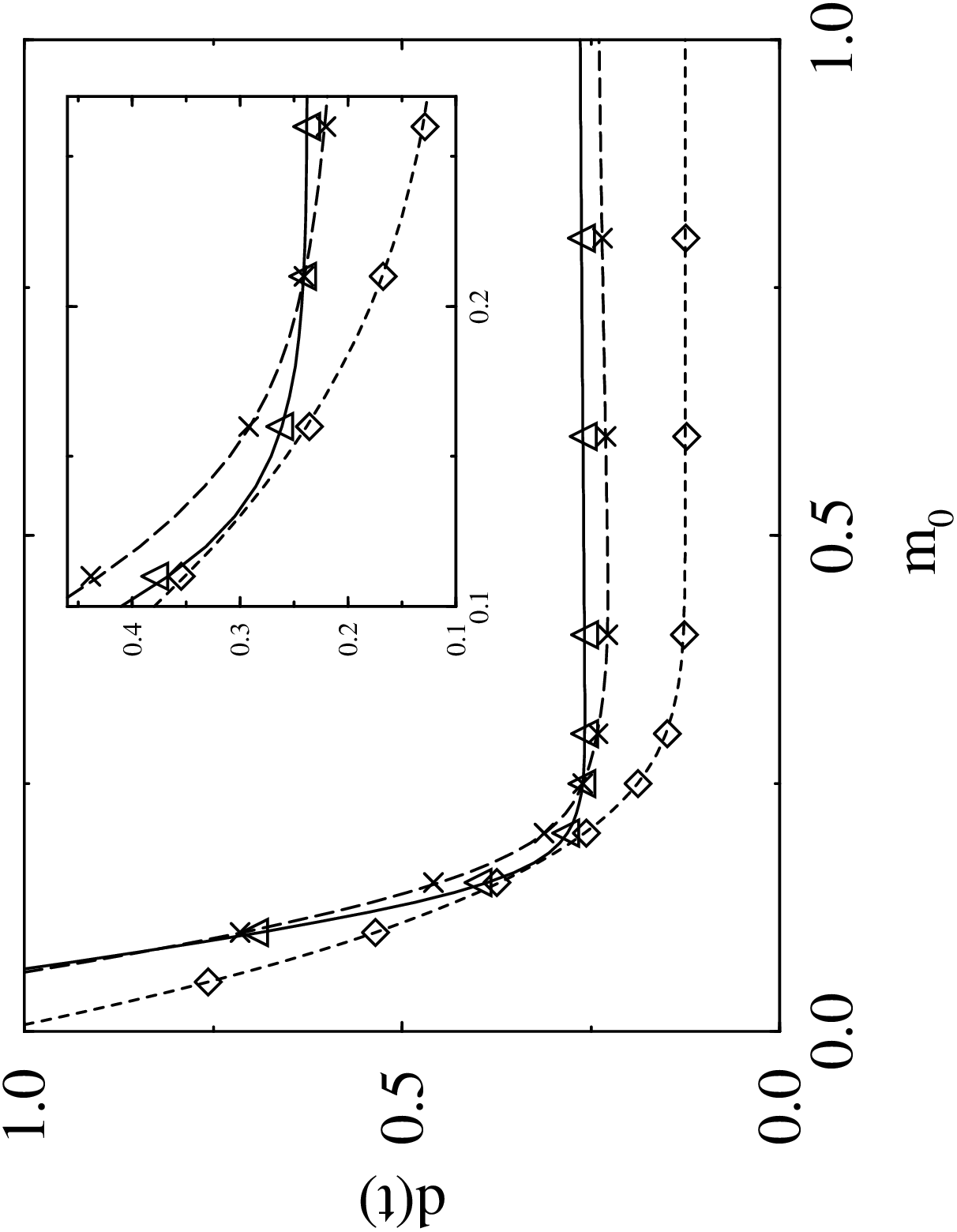}}}
\caption{
As in Fig. 2, for the network parameters $b=0.1, \alpha= 0.015, a_0=0.85$.}
\end{figure}

\end{document}